\documentclass[letterpaper,twocolumn,prl,aps,superscriptaddress,floatfix]{revtex4}
\pdfoutput=1
\usepackage{mathptmx}

\usepackage[latin9]{inputenc}
\usepackage{amsmath}
\usepackage{amssymb}
\usepackage{graphicx}
\usepackage{esint}

\makeatletter
\pdfpageheight\paperheight
\pdfpagewidth\paperwidth

\@ifundefined{textcolor}{}
{%
 \definecolor{BLACK}{gray}{0}
 \definecolor{WHITE}{gray}{1}
 \definecolor{RED}{rgb}{1,0,0}
 \definecolor{GREEN}{rgb}{0,1,0}
 \definecolor{BLUE}{rgb}{0,0,1}
 \definecolor{CYAN}{cmyk}{1,0,0,0}
 \definecolor{MAGENTA}{cmyk}{0,1,0,0}
 \definecolor{YELLOW}{cmyk}{0,0,1,0}
}
\usepackage{xcolor}
\usepackage{soul}

\newcommand{\ket}[1]{\ensuremath{\left|#1\right\rangle}}
\definecolor{blue}{rgb}{0,0,1}
\definecolor{red}{rgb}{1,0,0}
\definecolor{green}{rgb}{0,1,0}

\usepackage[unicode=true,pdfusetitle,
 bookmarks=true,bookmarksnumbered=false,bookmarksopen=false,
 breaklinks=false,pdfborder={0 0 0},pdfborderstyle={},backref=false,colorlinks=true]
 {hyperref}
\hypersetup{
 colorlinks,linkcolor=red,citecolor=blue}

\makeatother

\begin{document}
\draft
\title{Manipulating complex hybrid entanglement and testing multipartite
Bell inequalities in a superconducting circuit}

\author{Y.~Ma}
\thanks{These two authors contributed equally to this work.}
\affiliation{Center for Quantum Information, Institute for Interdisciplinary Information Sciences, Tsinghua University, Beijing 100084, China}

\author{X.~Pan}
\thanks{These two authors contributed equally to this work.}
\affiliation{Center for Quantum Information, Institute for Interdisciplinary Information Sciences, Tsinghua University, Beijing 100084, China}

\author {W.~Cai}
\affiliation{Center for Quantum Information, Institute for Interdisciplinary Information Sciences, Tsinghua University, Beijing 100084, China}

\author {X.~Mu}
\affiliation{Center for Quantum Information, Institute for Interdisciplinary Information Sciences, Tsinghua University, Beijing 100084, China}

\author {Y.~Xu}
\affiliation{Center for Quantum Information, Institute for Interdisciplinary Information Sciences, Tsinghua University, Beijing 100084, China}

\author {L.~Hu}
\affiliation{Center for Quantum Information, Institute for Interdisciplinary Information Sciences, Tsinghua University, Beijing 100084, China}

\author {W.~Wang}
\affiliation{Center for Quantum Information, Institute for Interdisciplinary Information Sciences, Tsinghua University, Beijing 100084, China}

\author {H.~Wang}
\affiliation{Center for Quantum Information, Institute for Interdisciplinary Information Sciences, Tsinghua University, Beijing 100084, China}

\author {Y.~P.~Song}
\affiliation{Center for Quantum Information, Institute for Interdisciplinary Information Sciences, Tsinghua University, Beijing 100084, China}

\author{Zhen-Biao~Yang}
\thanks{E-mail:zbyang@fzu.edu.cn}
\affiliation{Fujian Key Laboratory of Quantum Information and Quantum Optics,\\
College of Physics and Information Engineering, Fuzhou University, Fuzhou,\\
Fujian 350116, China}

\author{Shi-Biao~Zheng}
\thanks{E-mail: t96034@fzu.edu.cn}
\affiliation{Fujian Key Laboratory of Quantum Information and Quantum Optics,\\
College of Physics and Information Engineering, Fuzhou University, Fuzhou,\\
Fujian 350116, China}

\author{L.~Sun}
\thanks{Email: luyansun@tsinghua.edu.cn}
\affiliation{Center for Quantum Information, Institute for Interdisciplinary Information Sciences, Tsinghua University, Beijing 100084, China}


\begin{abstract}
Quantum correlations in observables of multiple systems not only are of
fundamental interest, but also play a key role in quantum information
processing. As a signature of these correlations, the violation of Bell
inequalities has not been demonstrated with multipartite hybrid entanglement
involving both continuous and discrete variables. Here we create a five-partite entangled state with three superconducting transmon qubits and two photonic qubits, each encoded in the mesoscopic field of a microwave cavity. We reveal the quantum correlations among these distinct elements by joint Wigner tomography of the two cavity fields conditional on the detection of the qubits and by test of
a five-partite Bell inequality. The measured Bell signal is $8.381\pm0.038$, surpassing the bound of 8 for a four-partite entanglement imposed by quantum correlations by 10 standard deviations, demonstrating the genuine five-partite entanglement in a hybrid quantum system.
\end{abstract}

\maketitle
\vskip 0.5cm
\narrowtext


The ability of controllably entangling multiple quantum systems and individually detecting their states is of importance both from the fundamental viewpoint and for practical applications, e.g., quantum computation. Essentially, carrying out a quantum algorithm is physically equivalent to preparing and manipulating entanglement for many two-dimensional systems (qubits) in a prescribed manner, and then reading out their states~\cite{QCaQI,Rauss01}. Among various kinds of multipartite entangled states, the Greenber-Horne-Zeilinger (GHZ) states represent a typical example~\cite{Greeberger90}. These states are formed by two maximally distinct joint quantum states of three or more qubits, whose properties can exhibit strong quantum correlations that exclude any local realistic description of nature in an all-or-nothing
manner~\cite{Greeberger90} or by violation of multipartite Bell inequalities~\cite{Bell65,Clauser69,Merin90}. The ratio of the Bell's signal associated with a GHZ state to the bound allowed by local realism increases with the number of entangled qubits, indicating the larger the entanglement the stronger the nonclassical effect~\cite{Merin90}. Besides fundamental interest, such states are a key resource for quantum-based technologies, including concatenated error correcting codes~\cite{Knill2005}, quantum simulation~\cite{ZhongEmulating,SongDemonstration}, and Heisenberg-limited quantum metrology~\cite{LeibfriedToward}. So far, GHZ states have been demonstrated for 10 photons~\cite{WangExperimental} and 14 trapped ions~\cite{Monz201014}; with these two systems, experimental violations of multipartite Bell inequalities have also been reported~\cite{LanyonExperimental,Walther2005,WBGao2010}.

Circuit quantum electrodynamics (cQED) systems, with superconducting qubits
coupled to resonators, are ideal for complex entanglement manipulation and
quantum information processing~\cite{You2003,Blais2004,Wallraff2004,You2011,Devoret2013,Gu2017Microwave}. Based on such systems, a variety of entangled states have been produced, including multiphoton NOON states for two resonators~\cite{HWang2011,YYGao}, two-mode cat states for mesoscopic fields stored in two cavities~\cite{CWang2016}, and multiqubit GHZ states~\cite{Paik2016,CSong2017,Song2019}. Using entangled states of two superconducting qubits coupled to a resonator, violations of the
Clauser-Horne-Shimony-Holt version of the Bell inequality have been
demonstrated~\cite{Ansmann2009,ChowPRA2010,Zhong2019}. Recently, this violation was detected between two encoded~cavity cat state qubits~\cite{CWang2016} and between a superconducting transmon qubit and an encoded cavity cat state qubit~\cite{Vlastakis2015}; in these experiments, the Bell test was used to characterize two-partite entanglement other than to vindicate quantum nonlocality due to the lack of independent measurements of the two entangled constituents.

In this letter we experimentally produce GHZ states for three superconducting
transmon qubits and two encoded cavity cat qubits in a three-dimensional
cQED architecture~\cite{Paik2011}. The entanglement among the three transmon qubits and the mesoscopic fields stored in the two cavities are generated by using the qubit-state-dependent cavity phase shifts and cavity-photon-number-dependent qubit rotations, enabled by the dispersive couplings between the cavities and the corresponding transmon qubits. As far as we know, this represents the largest hybrid entangled state reported so far; previously, hybrid entanglement was
restricted to one discrete variable and two continuous variables~\cite{CWang2016}. We characterize the multipartite entanglement by measuring the joint Wigner
functions for the two cavity fields conditional on outcomes of joint qubit
detection and by testing the multipartite Bell inequality. We measure a Bell
signal of $8.381\pm 0.038$, surpassing the maximum value of 8 allowed by quantum mechanics for a 4-partite quantum system. The results demonstrate a good control over complex three-dimensional cQED systems, which represent a promising platform for fault-tolerant quantum computation~\cite{Ma2019,Reinhold2019}.

\begin{figure}
\includegraphics[scale=1]{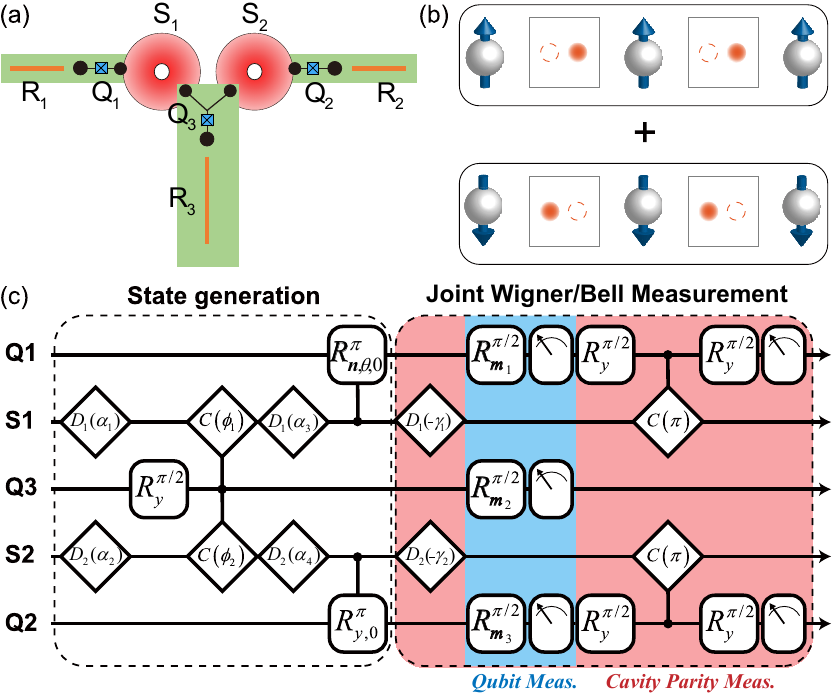}
\caption{(Color online) Device sketch and pulse sequence. (a) Device sketch.
The device involves three superconducting transmon qubits, labelled from $Q_{1}$ to $Q_{3}$, with $Q_{1}$ ($Q_{2}$) dispersively coupled to the first
(second) storage cavity, denoted as $S_{1}$ ($S_{2}$), and $Q_{3}$ coupled
to both storage cavities. Each qubit is independently coupled to a readout
cavity. The storage cavities are used for storing the photonic fields, while
the readout cavities for detecting the qubit states. (b) Schematic of hybrid entanglement containing three discrete variables and two continuous variables. (c) Experimental pulse sequence to create and characterize the entangled state in (b). The procedure consists of three parts. 1) Generation of the multipartite entangled state for the three transmon qubits and the two storage cavities, realized by a
sequence of operations, including initialization of the system to the ground
state, a pair of phase-space displacements $D_{1}(\alpha _{1}=1.782)$ and $D_{2}(\alpha _{2}=1.782)$ on the two cavity fields, rotation $R_{y}^{\pi /2}$ of $Q_{3}$, $Q_{3}$-state-dependent cavity phase shifts, a second pair of cavities displacements $D_{1}(\alpha_{3})$ and $D_{2}(\alpha_{4})$, and $\pi$ rotations $R_{\mathbf{n},\theta,0}^{\pi}$ and $R_{y,0}^{\pi}$ to $Q_{1}$ and $Q_{2}$ conditional on the vacuum states of $S_1$ and $S_2$, respectively. The angle ($\theta$) between the rotation axis of $R_{\mathbf{n},\theta,0}^{\pi}$ and $\mathbf{n}$-axis is variable. 2) Measurement of the qubit observables with the appropriate $\pi/2$ pre-rotations around the $\mathbf{m}$-axis on the equatorial plane $R_{\mathbf{m}_j}^{\pi/2}$ before the state readout in the basis \{$\ket{g_{j}}, \ket{e_{j}}$\}. $R_{\mathbf{m}_j}^{\pi/2}=R_{-y}^{\pi /2}$ for the joint Wigner measurement and $R_{\mathbf{m}_j}^{\pi/2}=R_{-y}^{\pi /2}$ or $R_{\pm x}^{\pi /2}$ for the Bell signal measurement. 3) Displaced photon-number parity detection for each cavity, realized by performing a phase-space displacement $D_{j}(-\gamma_{j})$ and then sandwiching a conditional cavity $\pi $-phase shift between two qubit rotations $R_{y}^{\pi /2}$ on the corresponding qubit. $D_{j}(-\gamma_{j})$ is used for conditional Wigner tomography and for $\sigma_{x}$ measurement of the corresponding cavity cat state qubit. For $\sigma_{y}$ measurement of each cavity, the displacement $D_{j}(-\gamma_{j})$ is replaced by the combination of two perpendicular displacements (see the main text).}
\label{Fig:fig1}
\end{figure}

Our experiment is performed with a cQED architecture involving three superconducting transmon qubits, two storage cavities serving for storing photonic fields, and three readout cavities, each of which
is dispersively coupled to one qubit for measuring the qubit state, as
sketched in Fig.~\ref{Fig:fig1}(a). The detailed device geometry can be found in Ref.~\cite{Xu2019} and the system parameters are listed in Supplementary Materials~\cite{Supplement}. As the readout cavities remain in the vacuum state and do not affect the quantum state of the qubits and storage cavities during the entanglement production, we can ignore the state of each readout cavity and will refer the corresponding storage cavity to as ``cavity''. Qubit $Q_1$ ($Q_2$) is dispersively coupled to cavity $S_1$ ($S_2$), while qubit $Q_3$ is commonly coupled to both cavities. In the interaction picture, the Hamiltonian for the total system is
\begin{eqnarray}
H_{I} &=&-a_{1}^{\dagger }a_{1}\left( \chi_{11}\left\vert e\right\rangle
_{1}\left\langle e\right\vert +\chi_{13}\left\vert e\right\rangle
_{3}\left\langle e\right\vert \right) \notag\\
&&-a_{2}^{\dagger }a_{2}\left( \chi_{22}\left\vert e\right\rangle
_{2}\left\langle e\right\vert +\chi_{23}\left\vert e\right\rangle
_{3}\left\langle e\right\vert \right) ,
\label{equ:Ham}
\end{eqnarray}%
where $a_{j}^{\dagger}$ and $a_{j}$ are the creation and annihilation
operators of the photonic field in cavity $S_j$ $(j=1,2)$, $\ket{e}_{k}$ denotes the excited state of qubit $k$ $(k=1,2,3)$, and $\chi_{jk}$ is the frequency shift of cavity $S_j$ conditional on the qubit state $\ket{e}_k$ due to the dispersive coupling.

Based on the above Hamiltonian [Eq.~(\ref{equ:Ham})] and the conditional operations, we can generate a hybrid entangled state as in Fig.~\ref{Fig:fig1}(b). Figure~\ref{Fig:fig1}(c) shows the pulse sequence for the creation and the following characterization of this state. The experiment starts with
initializing each qubit to the ground state $\left\vert g\right\rangle $ and
each cavity to vacuum state $\left\vert 0\right\rangle $. After this
initialization, we apply a microwave pulse to each cavity to produce a
displacement operation $D_{j}(\alpha _{j})$ in phase space, translating the
cavity from the vacuum state $\left\vert 0\right\rangle _{j}$ to a coherent
state $\left\vert \alpha _{j}\right\rangle _{j}$, with $\alpha _{j}$ being
the complex amplitude of the phase-space displacement. The subsequent
rotation $R_{y}^{\pi /2}$ (a $\pi/2$ rotation around the $y$-axis) on qubit $Q_3$, achieved by the application of a driving pulse, transforming $\left\vert g\right\rangle _{3}$ to $\left(\left\vert g\right\rangle _{3}+\left\vert e\right\rangle _{3}\right) /\sqrt{2}$. After an interaction time $\tau $, the dispersive interaction of qubit $Q_3$
with each cavity leads to a conditional phase shift~\cite{CWang2016,Vlastakis2015,Vlastakis2013,Sun2014,Liu2017}, evolving the
system to the state
\begin{eqnarray}
\left\vert g\right\rangle _{1}\left\vert g\right\rangle _{2}\left(
\left\vert g\right\rangle _{3}\left\vert \alpha _{1}\right\rangle
_{1}\left\vert \alpha _{2}\right\rangle _{2}+\left\vert e\right\rangle
_{3}\left\vert \alpha _{1}e^{i\phi _{1}}\right\rangle _{1}\left\vert \alpha
_{2}e^{i\phi _{2}}\right\rangle _{2}\right) /\sqrt{2},
\label{equ:Disstate1}
\end{eqnarray}
where $\phi _{j}=\chi_{j3}\tau$. A second displacement operation, $D_{1}(\alpha_3=-\alpha_{1})$, is then applied to cavity $S_1$, and $D_{2}(\alpha_4=-\alpha_{2}e^{i\phi _{2}})$ to cavity $S_2$, resulting in the state
\begin{eqnarray}
\frac{1}{\sqrt{2}}\left\vert g\right\rangle _{1}\left\vert g\right\rangle _{2}\left(
e^{-i\left\vert \alpha _{2}\right\vert ^{2}\sin \phi _{2}}\left\vert
g\right\rangle _{3}\left\vert 0\right\rangle _{1}\left\vert \alpha
_{2}-\alpha _{2}e^{i\phi _{2}}\right\rangle _{2}\right.\notag\\
+\left.e^{i\left\vert \alpha
_{1}\right\vert ^{2}\sin \phi _{1}}\left\vert e\right\rangle _{3}\left\vert
\alpha _{1}e^{i\phi _{1}}-\alpha _{1}\right\rangle _{1}\left\vert
0\right\rangle _{2}\right).
\label{equ:Disstate2}
\end{eqnarray}

After this operation, we perform a $\pi $ rotation to each qubit conditional
on the vacuum state of the corresponding cavity~\cite{HWang2011,Vlastakis2013,Liu2017}, realized by a pulse resonant with the qubit transition associated with the cavity's vacuum state $\ket{0}$. The conditional $\pi$ rotation on $Q_1$, denoted as $R_{\mathbf{n},\theta,0}^{\pi}$, is around an axis with an adjustable angle $\theta$ relative to $\mathbf{n}$-axis (reference axis) which has an angle $\pi /2+\left\vert \alpha _{1}\right\vert ^{2}\sin \phi _{1}+\left\vert \alpha _{2}\right\vert ^{2}\sin \phi _{2}$ to the $x$-axis on the equatorial plane, while that on $Q_2$ ($R_{y,0}^{\pi}$) is around the $y$-axis. After these conditional rotations, the total system evolves to
\begin{eqnarray}
\left( \left\vert e\right\rangle _{1}\left\vert g\right\rangle
_{2}\left\vert g\right\rangle _{3}\left\vert 0\right\rangle _{1}\left\vert
\beta _{2}\right\rangle _{2}+e^{-i\theta}\left\vert g\right\rangle _{1}\left\vert
e\right\rangle _{2}\left\vert e\right\rangle _{3}\left\vert \beta
_{1}\right\rangle _{1}\left\vert 0\right\rangle _{2}\right) /\sqrt{2},
\label{equ:GHZstate}
\end{eqnarray}
where $\beta _{1}=$ $\left( \alpha _{1}e^{i\phi _{1}}-\alpha _{1}\right)
e^{i\chi _{13}\tau ^{^{\prime }}}$ and $\beta _{2}=$ $\alpha _{2}-$ $\alpha
_{2}e^{i\phi _{2}}$, with $\tau ^{^{\prime}}$ being the duration of these
conditional rotations. When $\left\vert \beta _{j}\right\vert ^{2}\gg 1$, $%
\left\vert \beta _{j}\right\rangle _{j}$ and $\left\vert 0\right\rangle _{j}$
are approximately orthogonal and can be considered as the two logic states of a
qubit. With this encoding, the state of Eq.~(\ref{equ:GHZstate}) represents a 5-partite GHZ state involving three transmon qubits and two cavity cat state qubits. In our experiment, $\beta_{1}=-2.7-0.2i$ and $\beta _{2}=0.8+2.3i$, corresponding to $\left\vert _{1}\left\langle \beta_{1}\right\vert \left. 0\right\rangle _{1}\right\vert ^{2}\simeq 6.6\times 10^{-4}$ and $\left\vert _{2}\left\langle \beta _{2}\right\vert \left. 0\right\rangle
_{2}\right\vert ^{2}\simeq 2.5\times 10^{-3}$.

\begin{figure}
\includegraphics[scale=1]{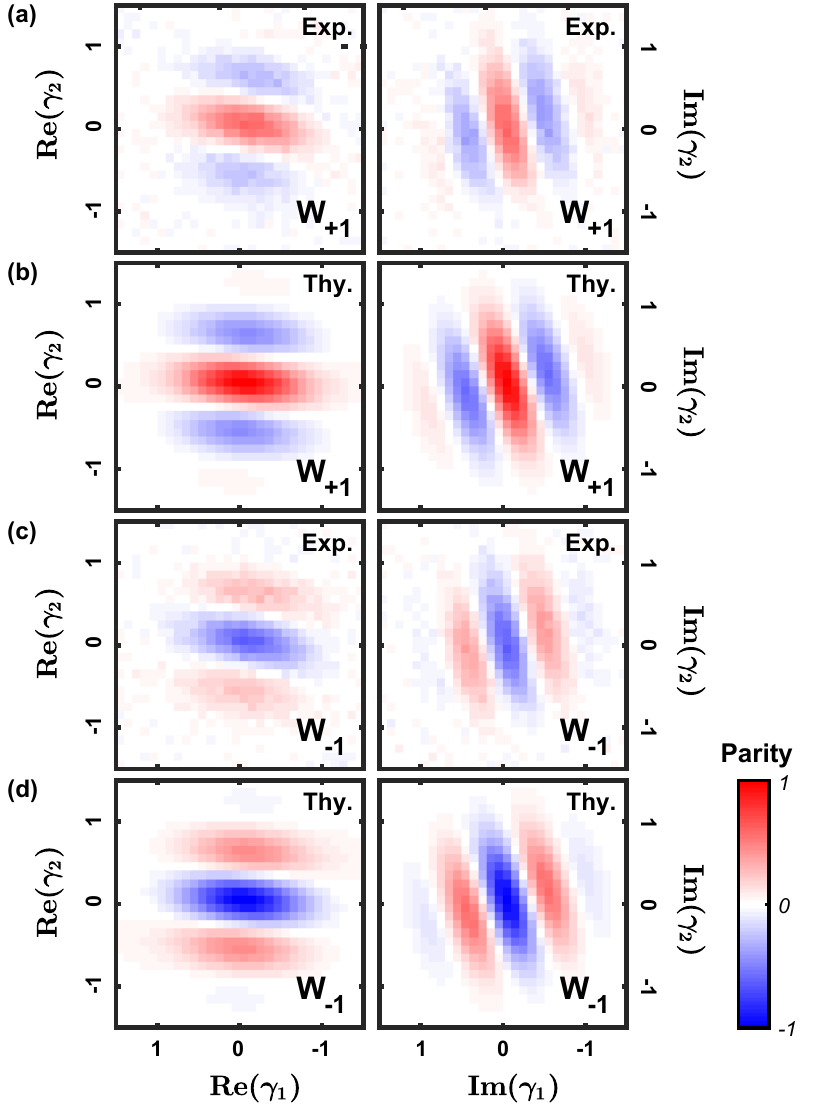}
\caption{(Color online) Conditional two-mode Wigner tomography for $S_1$ and $S_2$ after generation of the GHZ state of Eq.~(\ref{equ:GHZstate}) with $\theta=0$. Two-dimensional plane-cuts along the Re($\gamma_1$)-Re($\gamma_2$) axes and Im($\gamma_1$)-Im($\gamma_2$) axes are shown in the left and right panels, respectively. We displace each mode with $-\beta_1/2$ and $-\beta_2/2$ respectively so that the center of fringes is the origin. (a) Cuts of the measured conditional Wigner function, $W_1(\gamma_1,\gamma_2)$, defined as the joint Wigner functions for the two cavity fields conditional on $X_1X_2X_3=1$. (b) The expected $W_1(\gamma_1,\gamma_2)$, calculated from the ideal state of Eq.~(\ref{equ:GHZstate}) with $\beta_1=-2.7-0.2i$ and $\beta_2=0.8+2.3i$. (c) Cuts of the measured $W_{-1}(\gamma_1,\gamma_2) $ conditional on $X_1X_2X_3=-1$. (d) The expected $W_{-1}(\gamma_1,\gamma_2)$, also calculated from the ideal state as (a).}
\label{Fig:fig2}
\end{figure}

To verify the entanglement among the transmon qubits and the cavity fields after the GHZ state generation, we measure the conditional two-mode Wigner functions $W_{1}(\gamma_{1},\gamma_{2})$ and $W_{-1}(\gamma_{1},\gamma_{2})$ [see Fig.~\ref{Fig:fig1}(c)], which are defined as the joint Wigner functions for the two cavity fields conditional on $X_{1}X_{2}X_{3}=1$ and $-1$, respectively. Here $X_{1}X_{2}X_{3}$ represents the value of the observable $\sigma _{x}^{1}\sigma _{x}^{2}\sigma _{x}^{3}$, with $\sigma_{x}^{j}$ being the $x$-component of the Pauli operator associated with the $j$-th transmon qubit. The observable $\sigma _{x}^{k}$ is measured by performing the rotation $R_{-y}^{\pi /2}$ on the $k$-th qubit and then reading out its state in the basis $\left\{ \left\vert g\right\rangle_{k},\left\vert e\right\rangle _{k}\right\} $; the outcomes $\left\vert g\right\rangle _{k}$ and $\left\vert e\right\rangle _{k}$ correspond to $X_{k}=1$ and $-1$, respectively. After the measurement of $X_{1}$ ($X_{2}$),
transmon qubit $Q_1$ ($Q_2$) is used to measure the displaced photon-number parity
of cavity $S_1$ ($S_2$), achieved by sandwiching a conditional cavity $\pi $-phase
shift between two qubit rotations $R_{y}^{\pi /2}$ after the corresponding
cavity displacement~\cite{CWang2016,Vlastakis2015,Vlastakis2013,Sun2014,Liu2017,WWang2017,Bertet2002}. The joint displaced photon-number parity of
two cavity fields is directly related with the two-mode Wigner function as
\begin{eqnarray}
W(\gamma_{1},\gamma_{2})=\frac{4}{\pi ^{2}}\left\langle P_{1,\gamma_{1}}P_{2,\gamma_{2}}\right\rangle \text{,}
\label{equ:JointWigner}
\end{eqnarray}%
where $P_{j,\gamma_{j}}=D_{j}(\gamma_{j})P_{j}D_{j}^{\dagger }(\gamma_{j})$,
with $P_{j}$ denoting the photon-number parity operator for cavity $S_j$. The
two-mode Wigner function is a function in a four-dimensional space spanned
by \{Re($\gamma_{1}$), Re($\gamma_{2}$), Im($\gamma_{1}$), Im($\gamma_{2}$)\}.
We measure $W(\gamma_{1},\gamma_{2})$ for both $X_{1}X_{2}X_{3}=1$ and $-1$.

The two-dimensional plane-cuts of the measured conditional Wigner function $W_{1}(\gamma_{1},\gamma_{2})$ and the ideal results are displayed in Figs.~\ref{Fig:fig2}(a) and \ref{Fig:fig2}(b) respectively, while those associated with $W_{-1}(\gamma_{1},\gamma_{2})$ are shown in
Figs.~\ref{Fig:fig2}(c) and \ref{Fig:fig2}(d). Here the relative phase between the two components of the produced GHZ state is $\theta=0$. Cuts along the Re($\gamma_{1}$)-Re($\gamma_{2}$) axes and Im($\gamma_{1}$)-Im($\gamma_{2}$) axes are shown in the left and right panels, respectively. As expected, both $W_{1}(\gamma_{1},\gamma_{2})$ and $W_{-1}(\gamma_{1},\gamma_{2})$ exhibit strong two-mode quantum interference features, evidenced by the fringes on the Re($\gamma_{1}$)-Re($\gamma_{2}$) and Im($\gamma_{1}$)-Im($\gamma_{2}$) planes~\cite{CWang2016,Milman2005}. The
complementarity between the interference patterns of $W_{1}(\gamma_{1},\gamma_{2})$ and $W_{-1}(\gamma_{1},\gamma_{2})$ reveals the entanglement between the
transmon qubits and the cavity fields.


\begin{figure}
\includegraphics[scale=1]{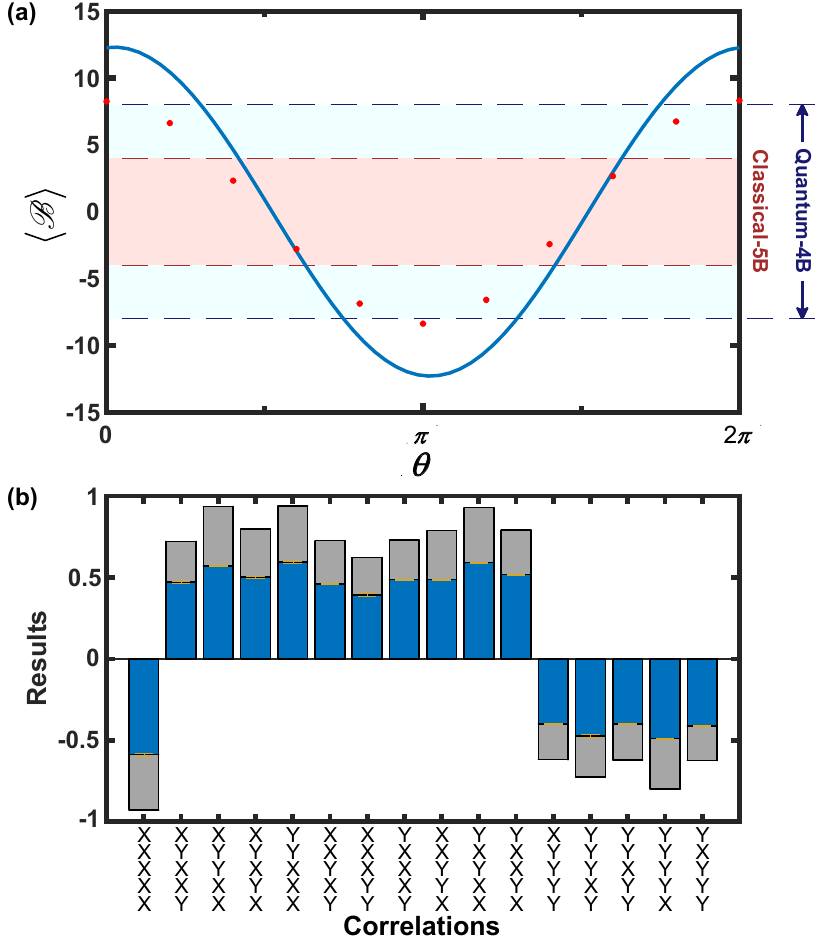}
\caption{(Color online) Bell signal measurement. (a) Measured expectation value of the Bell operator $\left\langle {\cal B}\right\rangle$ as a function of the relative phase ($\theta$) of the produced GHZ state. An oscillation is observed as predicted. Experiment data are marked with red. Their standard deviations are obtained from 10 repeated measurements when the system is stable and are smaller than marker sizes. Simulated results based on QuTiP in Python~\cite{Johansson2012,Johansson2013} are shown as the blue curve. In the simulation, we neglect the system decoherence in the GHZ state generation and calculate the expectation value of the Bell operator of Eq.~(\ref{equ:Belloper}) with thus obtained GHZ state. The maximum of the measured Bell signals $|\left\langle {\cal B}\right\rangle| =8.381\pm0.038$ not only exceeds the bound of 4 allowed by classical models (red region), but also above the bound of 8 for a 4-partite entangled quantum system (light blue region), demonstrating the genuine 5-partite entanglement in a hybrid system. (b) Correlations combining the maximum of the measured Bell signals in (a). Blue bars are experimental data and grey ones represent simulated results. Error bars are from repeated measurements. The measured results are lower due to the imperfections in the experiment (see the main text), but the overall distribution is consistent with simulation.}
\label{Fig:fig3}
\end{figure}

The 5-partite entanglement can be further revealed by the multipartite Bell
inequality, proposed by Mermin~\cite{Merin90}. For clarity, we rewrite this state as
\begin{eqnarray}
\left( \left\vert \uparrow \right\rangle _{1}\left\vert \uparrow
\right\rangle _{2}\left\vert \uparrow \right\rangle _{3}\left\vert \uparrow
\right\rangle _{4}\left\vert \uparrow \right\rangle _{5}+\left\vert
\downarrow \right\rangle _{1}\left\vert \downarrow \right\rangle
_{2}\left\vert \downarrow \right\rangle _{3}\left\vert \downarrow
\right\rangle _{4}\left\vert \downarrow \right\rangle _{5}\right) /\sqrt{2},
\label{equ:Bellsymb}
\end{eqnarray}
where $\left\vert \uparrow \right\rangle _{1}\equiv $ $\left\vert
e\right\rangle _{1}$ and $\left\vert \downarrow \right\rangle _{1}\equiv
\left\vert g\right\rangle _{1}$; $\left\vert \uparrow \right\rangle
_{k}\equiv $ $\left\vert g\right\rangle _{k}$ and $\left\vert \downarrow
\right\rangle \equiv \left\vert e\right\rangle _{k}$ for $k=2$, $3$; $%
\left\vert \uparrow \right\rangle _{4}\equiv \left\vert 0\right\rangle _{1}$
and $\left\vert \downarrow \right\rangle _{4}$ $\equiv \left\vert \beta
_{1}\right\rangle _{1}$; $\left\vert \uparrow \right\rangle
_{5}\equiv \left\vert \beta _{2}\right\rangle _{2}$, $\left\vert \downarrow
\right\rangle _{5}$ $\equiv \left\vert 0\right\rangle _{2}$. The Bell
operator for this hybrid system is defined as
\begin{eqnarray}
{\cal B} &=&\sigma _{x}^{1}\sigma _{x}^{2}\sigma _{x}^{3}\sigma
_{x}^{4}\sigma _{x}^{5}-\sum_{l}P_{l}\left( \sigma _{x}^{1}\sigma
_{x}^{2}\sigma _{x}^{3}\sigma _{y}^{4}\sigma _{y}^{5}\right)  \notag\\
&&\ +\sum_{l}P_{l}\left( \sigma _{x}^{1}\sigma _{y}^{2}\sigma _{y}^{3}\sigma
_{y}^{4}\sigma _{y}^{5}\right),
\label{equ:Belloper}
\end{eqnarray}
where $\{P_{l}\}$ denotes the set of all distinct permutations of the
subscripts that give distinct products. The measurement sequence is also shown in Fig.~\ref{Fig:fig1}(c). As has been shown, for $k\leq 3$ the
observables $\sigma_{x}^{k}$ is measured by performing the rotation $R_{-y}^{\pi /2}$ before state readout of the corresponding qubit. $\sigma
_{y}^{1}$ is measured by performing the rotation $R_{-x}^{\pi /2}$ before the
state readout; for the measurement of $\sigma _{y}^{2}$ and $\sigma _{y}^{3}$, $R_{-x}^{\pi /2}$ is replaced by $R_{x}^{\pi /2}$. For each of these
measurements, the value of $1$ ($-1$) is assigned to the corresponding
observable when the outcome is $\left\vert g\right\rangle _{k}$ ($\left\vert
e\right\rangle _{k}$). On the other hand, for $k=4$ and $5$, $\sigma_{x}^{k}$ and
$\sigma _{y}^{k}$ are measured by mapping them to displaced photon-number
parity observables of the $j$-th cavity with $j=k-3$~\cite{Vlastakis2015}: $\sigma _{x}^{k}$ corresponds to $D_{j}(\gamma_{j})P_{j}D_{j}^{\dagger}(\gamma_{j})$ with $D_{j}(\gamma_{j})=D_{j}(\beta_{j}/2)$; while $\sigma_{y}^k$ is approximated by $D_{j}(\gamma_{j})P_{j}D_{j}^{\dagger}(\gamma_{j})$ with $D_{j}(\gamma_{j})=D_{j}(\beta_{j}/2)D_{j}(-i\epsilon_{j}\pi /4\beta_{j})$, where $\epsilon _{j}=-1$ and $1$ for $j=1$ and $2$, respectively. The even and odd
parities of the $j$-th cavity after the corresponding displacement operation
correspond to the values $1$ and $-1$ for $X_{j+3}$ or $Y_{j+3}$,
respectively. The Bell operator involves 16 terms. The experiment is
repeated 10,000 times for the measurement of the correlation corresponding to
each term, which is obtained by averaging over all of the experimental
outcomes. All the correlation values are combined to obtain the
corresponding Bell signal according to the expansion of Eq.~(\ref{equ:Belloper}).

An ideal 5-qubit GHZ state is the eigenstate of each term of ${\cal B}$ with
the eigenvalue $1$, so that the ideal value of $\left\langle {\cal B}\right\rangle$ is $16$. The measured expectation value of the Bell operator, $\left\langle {\cal B}\right\rangle$, as a function of the relative phase ($\theta$) of the produced GHZ state is shown in Fig.~\ref{Fig:fig3}(a). Due to the large Hilbert space spanned by this 5-partite hybrid system, exact numerical simulations are difficult. Hence, we simplify the simulation by neglecting the system decoherence in the GHZ state generation and calculating the expectation value of the Bell operator of Eq.~(\ref{equ:Belloper}) with the simulated GHZ state. The simulated result is shown as the solid line. The correlation associated with each term of ${\cal B}$ at the maximum ($\theta=\pi$) is plotted in Fig.~\ref{Fig:fig3}(b). Blue bars are experimental data and grey ones correspond to the simulated results.

The simulated correlations deviate from 1 and give the maximum value of Bell signal of only $|\left\langle {\cal B}\right\rangle|=12.29$. This is mainly due to the following two factors. The first one is the Kerr effects which deform the coherent state components of the cavities and cause an imperfection of the conditional qubit rotations, but are not included in the Hamiltonian of Eq.~(\ref{equ:Ham}). The second one is the discrepancy between $\sigma_y$ for the photonic qubits and the corresponding displaced parity operators.

The measured values are even lower because of: 1) The qubit decoherence and cavity photon losses during the state generation and the following characterization; 2) the large cavity field amplitudes that cause the parity measurement following the displacements $D_{j}(-\gamma_j)$ to deviate from the expected cavity observables; 3) the readout errors of the qubit states, to which the Bell signal is very sensitive. This is due to the fact that once a qubit readout error occurs, the sign of output value of the corresponding measured term is changed. Nevertheless, the resulting measured Bell signal is $|\left\langle {\cal B}\right\rangle|=8.381\pm0.038$ at the maximum, in good agreement with expectation after considering the above imperfections (Supplementary Materials~\cite{Supplement}). This value not only exceeds the bound of 4 allowed by classical models, but also is above the bound of 8 for a 4-partite entangled quantum system, demonstrating the genuine 5-partite entanglement in a hybrid system.


We have experimentally demonstrated controlled entanglement manipulation and
characterization in a cQED quantum processor involving three
superconducting transmon qubits and two cavities storing mesoscopic fields.
We deterministically entangled all these elements and verified their quantum
correlations by reconstructing the joint Wigner functions of the two
cavities conditional on the detection of the states of the qubits. We
further measure the 5-partite Bell signal of $8.381\pm0.038$, exceeding the maximum value of 8 for a 4-partite entangled state by 10 standard deviations. Apart from fundamental interest, our experiment serves as a demonstration of good control over a quantum circuit, which is important for solid-state quantum computation. For larger systems with qubit-cavity chain geometry, the multipartite GHZ states can be prepared in a sequential way based on the same scheme presented in this work.

This work was supported by the National Natural Science Foundation of China
under Grant Nos. 11674060, 11874114, and 11925404, and National Key Research and Development Program of China under Grant No. 2017YFA0304303.

%
%

%


\begin{thebibliography}{44}%
\makeatletter
\providecommand \@ifxundefined [1]{%
 \@ifx{#1\undefined}
}%
\providecommand \@ifnum [1]{%
 \ifnum #1\expandafter \@firstoftwo
 \else \expandafter \@secondoftwo
 \fi
}%
\providecommand \@ifx [1]{%
 \ifx #1\expandafter \@firstoftwo
 \else \expandafter \@secondoftwo
 \fi
}%
\providecommand \natexlab [1]{#1}%
\providecommand \enquote  [1]{``#1''}%
\providecommand \bibnamefont  [1]{#1}%
\providecommand \bibfnamefont [1]{#1}%
\providecommand \citenamefont [1]{#1}%
\providecommand \href@noop [0]{\@secondoftwo}%
\providecommand \href [0]{\begingroup \@sanitize@url \@href}%
\providecommand \@href[1]{\@@startlink{#1}\@@href}%
\providecommand \@@href[1]{\endgroup#1\@@endlink}%
\providecommand \@sanitize@url [0]{\catcode `\\12\catcode `\$12\catcode
  `\&12\catcode `\#12\catcode `\^12\catcode `\_12\catcode `\%12\relax}%
\providecommand \@@startlink[1]{}%
\providecommand \@@endlink[0]{}%
\providecommand \url  [0]{\begingroup\@sanitize@url \@url }%
\providecommand \@url [1]{\endgroup\@href {#1}{\urlprefix }}%
\providecommand \urlprefix  [0]{URL }%
\providecommand \Eprint [0]{\href }%
\providecommand \doibase [0]{http://dx.doi.org/}%
\providecommand \selectlanguage [0]{\@gobble}%
\providecommand \bibinfo  [0]{\@secondoftwo}%
\providecommand \bibfield  [0]{\@secondoftwo}%
\providecommand \translation [1]{[#1]}%
\providecommand \BibitemOpen [0]{}%
\providecommand \bibitemStop [0]{}%
\providecommand \bibitemNoStop [0]{.\EOS\space}%
\providecommand \EOS [0]{\spacefactor3000\relax}%
\providecommand \BibitemShut  [1]{\csname bibitem#1\endcsname}%
\let\auto@bib@innerbib\@empty
\bibitem [{\citenamefont {Nielsen}\ and\ \citenamefont {Chuang}(2000)}]{QCaQI}%
  \BibitemOpen
  \bibfield  {author} {\bibinfo {author} {\bibfnamefont {M.~A.}\ \bibnamefont
  {Nielsen}}\ and\ \bibinfo {author} {\bibfnamefont {I.~L.}\ \bibnamefont
  {Chuang}},\ }\href@noop {} {\emph {\bibinfo {title} {Quantum Computation and
  Quantum Information}}}\ (\bibinfo  {publisher} {Cambridge Univ. Press},\
  \bibinfo {year} {2000})\BibitemShut {NoStop}%
\bibitem [{\citenamefont {Raussendorf}\ and\ \citenamefont
  {Briegel}(2001)}]{Rauss01}%
  \BibitemOpen
  \bibfield  {author} {\bibinfo {author} {\bibfnamefont {R.}~\bibnamefont
  {Raussendorf}}\ and\ \bibinfo {author} {\bibfnamefont {H.~J.}\ \bibnamefont
  {Briegel}},\ }\bibfield  {title} {\enquote {\bibinfo {title} {A one-way
  quantum computer},}\ }\href
  {https://journals.aps.org/prl/abstract/10.1103/PhysRevLett.86.5188}
  {\bibfield  {journal} {\bibinfo  {journal} {Phys. Rev. Lett.}\ }\textbf
  {\bibinfo {volume} {86}},\ \bibinfo {pages} {5188} (\bibinfo {year}
  {2001})}\BibitemShut {NoStop}%
\bibitem [{\citenamefont {Greenberger}\ \emph {et~al.}(1990)\citenamefont
  {Greenberger}, \citenamefont {Horne}, \citenamefont {Shimony},\ and\
  \citenamefont {Zeilinger}}]{Greeberger90}%
  \BibitemOpen
  \bibfield  {author} {\bibinfo {author} {\bibfnamefont {D.~M.}\ \bibnamefont
  {Greenberger}}, \bibinfo {author} {\bibfnamefont {M.~A.}\ \bibnamefont
  {Horne}}, \bibinfo {author} {\bibfnamefont {A.}~\bibnamefont {Shimony}}, \
  and\ \bibinfo {author} {\bibfnamefont {A.}~\bibnamefont {Zeilinger}},\
  }\bibfield  {title} {\enquote {\bibinfo {title} {Bell's theorem without
  inequalities},}\ }\href {https://doi.org/10.1119/1.16243} {\bibfield
  {journal} {\bibinfo  {journal} {Am. J. Phys.}\ }\textbf {\bibinfo {volume}
  {58}},\ \bibinfo {pages} {1131} (\bibinfo {year} {1990})}\BibitemShut
  {NoStop}%
\bibitem [{\citenamefont {Bell}(1965)}]{Bell65}%
  \BibitemOpen
  \bibfield  {author} {\bibinfo {author} {\bibfnamefont {J.~S.}\ \bibnamefont
  {Bell}},\ }\bibfield  {title} {\enquote {\bibinfo {title} {On the
  {Einstein-Podolsky-Rosen} paradox},}\ }\href
  {https://journals.aps.org/ppf/abstract/10.1103/PhysicsPhysiqueFizika.1.195}
  {\bibfield  {journal} {\bibinfo  {journal} {Physics}\ }\textbf {\bibinfo
  {volume} {1}},\ \bibinfo {pages} {195} (\bibinfo {year} {1965})}\BibitemShut
  {NoStop}%
\bibitem [{\citenamefont {Clauser}\ \emph {et~al.}(1969)\citenamefont
  {Clauser}, \citenamefont {Horne}, \citenamefont {Shimony},\ and\
  \citenamefont {Holt}}]{Clauser69}%
  \BibitemOpen
  \bibfield  {author} {\bibinfo {author} {\bibfnamefont {J.~F.}\ \bibnamefont
  {Clauser}}, \bibinfo {author} {\bibfnamefont {M.~A.}\ \bibnamefont {Horne}},
  \bibinfo {author} {\bibfnamefont {A.}~\bibnamefont {Shimony}}, \ and\
  \bibinfo {author} {\bibfnamefont {R.~A.}\ \bibnamefont {Holt}},\ }\bibfield
  {title} {\enquote {\bibinfo {title} {Proposed experiment to test local
  hidden-variable theories},}\ }\href
  {https://journals.aps.org/prl/abstract/10.1103/PhysRevLett.23.880} {\bibfield
   {journal} {\bibinfo  {journal} {Phys. Rev. Lett.}\ }\textbf {\bibinfo
  {volume} {23}},\ \bibinfo {pages} {880} (\bibinfo {year} {1969})}\BibitemShut
  {NoStop}%
\bibitem [{\citenamefont {Mermin}(1990)}]{Merin90}%
  \BibitemOpen
  \bibfield  {author} {\bibinfo {author} {\bibfnamefont {N.~D.}\ \bibnamefont
  {Mermin}},\ }\bibfield  {title} {\enquote {\bibinfo {title} {Extreme quantum
  entanglement in a superposition of macroscopically distinct states},}\ }\href
  {https://journals.aps.org/prl/abstract/10.1103/PhysRevLett.65.1838}
  {\bibfield  {journal} {\bibinfo  {journal} {Phys. Rev. Lett.}\ }\textbf
  {\bibinfo {volume} {65}},\ \bibinfo {pages} {1838} (\bibinfo {year}
  {1990})}\BibitemShut {NoStop}%
\bibitem [{\citenamefont {Knill}(2005)}]{Knill2005}%
  \BibitemOpen
  \bibfield  {author} {\bibinfo {author} {\bibfnamefont {E.}~\bibnamefont
  {Knill}},\ }\bibfield  {title} {\enquote {\bibinfo {title} {Quantum computing
  with realistically noisy devices},}\ }\href
  {https://www.nature.com/articles/nature03350} {\bibfield  {journal} {\bibinfo
   {journal} {Nature}\ }\textbf {\bibinfo {volume} {434}},\ \bibinfo {pages}
  {39} (\bibinfo {year} {2005})}\BibitemShut {NoStop}%
\bibitem [{\citenamefont {Zhong}\ \emph {et~al.}(2016)\citenamefont {Zhong},
  \citenamefont {Xu}, \citenamefont {Wang}, \citenamefont {Song}, \citenamefont
  {Guo}, \citenamefont {Liu}, \citenamefont {Xu}, \citenamefont {Xia},
  \citenamefont {Lu},\ and\ \citenamefont {Han}}]{ZhongEmulating}%
  \BibitemOpen
  \bibfield  {author} {\bibinfo {author} {\bibfnamefont {Y.~P.}\ \bibnamefont
  {Zhong}}, \bibinfo {author} {\bibfnamefont {D.}~\bibnamefont {Xu}}, \bibinfo
  {author} {\bibfnamefont {P.}~\bibnamefont {Wang}}, \bibinfo {author}
  {\bibfnamefont {C.}~\bibnamefont {Song}}, \bibinfo {author} {\bibfnamefont
  {Q.~J.}\ \bibnamefont {Guo}}, \bibinfo {author} {\bibfnamefont {W.~X.}\
  \bibnamefont {Liu}}, \bibinfo {author} {\bibfnamefont {K.}~\bibnamefont
  {Xu}}, \bibinfo {author} {\bibfnamefont {B.~X.}\ \bibnamefont {Xia}},
  \bibinfo {author} {\bibfnamefont {C.-Y.}\ \bibnamefont {Lu}}, \ and\ \bibinfo
  {author} {\bibfnamefont {S.~Y.}\ \bibnamefont {Han}},\ }\bibfield  {title}
  {\enquote {\bibinfo {title} {Emulating anyonic fractional statistical
  behavior in a superconducting quantum circuit},}\ }\href
  {https://journals.aps.org/prl/abstract/10.1103/PhysRevLett.117.110501}
  {\bibfield  {journal} {\bibinfo  {journal} {Phys. Rev. Lett.}\ }\textbf
  {\bibinfo {volume} {117}},\ \bibinfo {pages} {110501} (\bibinfo {year}
  {2016})}\BibitemShut {NoStop}%
\bibitem [{\citenamefont {Song}\ \emph {et~al.}(2018)\citenamefont {Song},
  \citenamefont {Xu}, \citenamefont {Zhang}, \citenamefont {Wang},
  \citenamefont {Guo}, \citenamefont {Liu}, \citenamefont {Xu}, \citenamefont
  {Deng}, \citenamefont {Huang}, \citenamefont {Zheng}, \citenamefont {Zheng},
  \citenamefont {Wang}, \citenamefont {Zhu}, \citenamefont {Lu},\ and\
  \citenamefont {Pan}}]{SongDemonstration}%
  \BibitemOpen
  \bibfield  {author} {\bibinfo {author} {\bibfnamefont {C.}~\bibnamefont
  {Song}}, \bibinfo {author} {\bibfnamefont {D.}~\bibnamefont {Xu}}, \bibinfo
  {author} {\bibfnamefont {P.}~\bibnamefont {Zhang}}, \bibinfo {author}
  {\bibfnamefont {J.}~\bibnamefont {Wang}}, \bibinfo {author} {\bibfnamefont
  {Q.}~\bibnamefont {Guo}}, \bibinfo {author} {\bibfnamefont {W.}~\bibnamefont
  {Liu}}, \bibinfo {author} {\bibfnamefont {K.}~\bibnamefont {Xu}}, \bibinfo
  {author} {\bibfnamefont {H.}~\bibnamefont {Deng}}, \bibinfo {author}
  {\bibfnamefont {K.}~\bibnamefont {Huang}}, \bibinfo {author} {\bibfnamefont
  {D.}~\bibnamefont {Zheng}}, \bibinfo {author} {\bibfnamefont {S.-B.}\
  \bibnamefont {Zheng}}, \bibinfo {author} {\bibfnamefont {H.}~\bibnamefont
  {Wang}}, \bibinfo {author} {\bibfnamefont {X.}~\bibnamefont {Zhu}}, \bibinfo
  {author} {\bibfnamefont {C.-Y.}\ \bibnamefont {Lu}}, \ and\ \bibinfo {author}
  {\bibfnamefont {J.-W.}\ \bibnamefont {Pan}},\ }\bibfield  {title} {\enquote
  {\bibinfo {title} {Demonstration of topological robustness of anyonic
  braiding statistics with a superconducting quantum circuit},}\ }\href
  {https://journals.aps.org/prl/abstract/10.1103/PhysRevLett.121.030502}
  {\bibfield  {journal} {\bibinfo  {journal} {Phys. Rev. Lett.}\ }\textbf
  {\bibinfo {volume} {121}},\ \bibinfo {pages} {030502} (\bibinfo {year}
  {2018})}\BibitemShut {NoStop}%
\bibitem [{\citenamefont {Leibfried}\ \emph {et~al.}(2004)\citenamefont
  {Leibfried}, \citenamefont {Barrett}, \citenamefont {Schaetz}, \citenamefont
  {Britton}, \citenamefont {Chiaverini}, \citenamefont {Itano}, \citenamefont
  {Jost}, \citenamefont {Langer},\ and\ \citenamefont
  {Wineland}}]{LeibfriedToward}%
  \BibitemOpen
  \bibfield  {author} {\bibinfo {author} {\bibfnamefont {D.}~\bibnamefont
  {Leibfried}}, \bibinfo {author} {\bibfnamefont {M.~D.}\ \bibnamefont
  {Barrett}}, \bibinfo {author} {\bibfnamefont {T.}~\bibnamefont {Schaetz}},
  \bibinfo {author} {\bibfnamefont {J.}~\bibnamefont {Britton}}, \bibinfo
  {author} {\bibfnamefont {J.}~\bibnamefont {Chiaverini}}, \bibinfo {author}
  {\bibfnamefont {W.~M.}\ \bibnamefont {Itano}}, \bibinfo {author}
  {\bibfnamefont {J.~D.}\ \bibnamefont {Jost}}, \bibinfo {author}
  {\bibfnamefont {C.}~\bibnamefont {Langer}}, \ and\ \bibinfo {author}
  {\bibfnamefont {D.~J.}\ \bibnamefont {Wineland}},\ }\bibfield  {title}
  {\enquote {\bibinfo {title} {Toward {Heisenberg}-limited spectroscopy with
  multiparticle entangled states},}\ }\href
  {https://science.sciencemag.org/content/304/5676/1476} {\bibfield  {journal}
  {\bibinfo  {journal} {Science}\ }\textbf {\bibinfo {volume} {304}},\ \bibinfo
  {pages} {1476} (\bibinfo {year} {2004})}\BibitemShut {NoStop}%
\bibitem [{\citenamefont {Wang}\ \emph
  {et~al.}(2016{\natexlab{a}})\citenamefont {Wang}, \citenamefont {Chen},
  \citenamefont {Li}, \citenamefont {Huang}, \citenamefont {Liu}, \citenamefont
  {Chen}, \citenamefont {Luo}, \citenamefont {Su}, \citenamefont {Wu},
  \citenamefont {Li}, \citenamefont {Lu}, \citenamefont {Hu}, \citenamefont
  {Jiang}, \citenamefont {Peng}, \citenamefont {Li}, \citenamefont {Liu},
  \citenamefont {Chen}, \citenamefont {Lu},\ and\ \citenamefont
  {Pan}}]{WangExperimental}%
  \BibitemOpen
  \bibfield  {author} {\bibinfo {author} {\bibfnamefont {X.-L.}\ \bibnamefont
  {Wang}}, \bibinfo {author} {\bibfnamefont {L.-K.}\ \bibnamefont {Chen}},
  \bibinfo {author} {\bibfnamefont {W.}~\bibnamefont {Li}}, \bibinfo {author}
  {\bibfnamefont {H.-L.}\ \bibnamefont {Huang}}, \bibinfo {author}
  {\bibfnamefont {C.}~\bibnamefont {Liu}}, \bibinfo {author} {\bibfnamefont
  {C.}~\bibnamefont {Chen}}, \bibinfo {author} {\bibfnamefont {Y.-H.}\
  \bibnamefont {Luo}}, \bibinfo {author} {\bibfnamefont {Z.-E.}\ \bibnamefont
  {Su}}, \bibinfo {author} {\bibfnamefont {D.}~\bibnamefont {Wu}}, \bibinfo
  {author} {\bibfnamefont {Z.-D.}\ \bibnamefont {Li}}, \bibinfo {author}
  {\bibfnamefont {H.}~\bibnamefont {Lu}}, \bibinfo {author} {\bibfnamefont
  {Y.}~\bibnamefont {Hu}}, \bibinfo {author} {\bibfnamefont {X.}~\bibnamefont
  {Jiang}}, \bibinfo {author} {\bibfnamefont {C.-Z.}\ \bibnamefont {Peng}},
  \bibinfo {author} {\bibfnamefont {L.}~\bibnamefont {Li}}, \bibinfo {author}
  {\bibfnamefont {N.-L.}\ \bibnamefont {Liu}}, \bibinfo {author} {\bibfnamefont
  {Y.-A.}\ \bibnamefont {Chen}}, \bibinfo {author} {\bibfnamefont {C.-Y.}\
  \bibnamefont {Lu}}, \ and\ \bibinfo {author} {\bibfnamefont {J.-W.}\
  \bibnamefont {Pan}},\ }\bibfield  {title} {\enquote {\bibinfo {title}
  {Experimental ten-photon entanglement},}\ }\href
  {https://journals.aps.org/prl/abstract/10.1103/PhysRevLett.117.210502}
  {\bibfield  {journal} {\bibinfo  {journal} {Phys. Rev. Lett.}\ }\textbf
  {\bibinfo {volume} {117}},\ \bibinfo {pages} {210502} (\bibinfo {year}
  {2016}{\natexlab{a}})}\BibitemShut {NoStop}%
\bibitem [{\citenamefont {Monz}\ \emph {et~al.}(2011)\citenamefont {Monz},
  \citenamefont {Schindler}, \citenamefont {Barreiro}, \citenamefont {Chwalla},
  \citenamefont {Nigg}, \citenamefont {Coish}, \citenamefont {Harlander},
  \citenamefont {H\"ansel}, \citenamefont {Hennrich},\ and\ \citenamefont
  {Blatt}}]{Monz201014}%
  \BibitemOpen
  \bibfield  {author} {\bibinfo {author} {\bibfnamefont {T.}~\bibnamefont
  {Monz}}, \bibinfo {author} {\bibfnamefont {P.}~\bibnamefont {Schindler}},
  \bibinfo {author} {\bibfnamefont {J.}~\bibnamefont {Barreiro}}, \bibinfo
  {author} {\bibfnamefont {M.}~\bibnamefont {Chwalla}}, \bibinfo {author}
  {\bibfnamefont {D.}~\bibnamefont {Nigg}}, \bibinfo {author} {\bibfnamefont
  {W.}~\bibnamefont {Coish}}, \bibinfo {author} {\bibfnamefont
  {M.}~\bibnamefont {Harlander}}, \bibinfo {author} {\bibfnamefont
  {W.}~\bibnamefont {H\"ansel}}, \bibinfo {author} {\bibfnamefont
  {M.}~\bibnamefont {Hennrich}}, \ and\ \bibinfo {author} {\bibfnamefont
  {R.}~\bibnamefont {Blatt}},\ }\bibfield  {title} {\enquote {\bibinfo {title}
  {14-qubit entanglement: creation and coherence},}\ }\href
  {https://journals.aps.org/prl/abstract/10.1103/PhysRevLett.106.130506}
  {\bibfield  {journal} {\bibinfo  {journal} {Phys. Rev. Lett.}\ }\textbf
  {\bibinfo {volume} {106}},\ \bibinfo {pages} {130506} (\bibinfo {year}
  {2011})}\BibitemShut {NoStop}%
\bibitem [{\citenamefont {Lanyon}\ \emph {et~al.}(2014)\citenamefont {Lanyon},
  \citenamefont {Zwerger}, \citenamefont {Jurcevic}, \citenamefont {Hempel},
  \citenamefont {D\"ur}, \citenamefont {Briegel}, \citenamefont {Blatt},\ and\
  \citenamefont {Roos}}]{LanyonExperimental}%
  \BibitemOpen
  \bibfield  {author} {\bibinfo {author} {\bibfnamefont {B.}~\bibnamefont
  {Lanyon}}, \bibinfo {author} {\bibfnamefont {M.}~\bibnamefont {Zwerger}},
  \bibinfo {author} {\bibfnamefont {P.}~\bibnamefont {Jurcevic}}, \bibinfo
  {author} {\bibfnamefont {C.}~\bibnamefont {Hempel}}, \bibinfo {author}
  {\bibfnamefont {W.}~\bibnamefont {D\"ur}}, \bibinfo {author} {\bibfnamefont
  {H.}~\bibnamefont {Briegel}}, \bibinfo {author} {\bibfnamefont
  {R.}~\bibnamefont {Blatt}}, \ and\ \bibinfo {author} {\bibfnamefont
  {C.}~\bibnamefont {Roos}},\ }\bibfield  {title} {\enquote {\bibinfo {title}
  {Experimental violation of multipartite {Bell} inequalities with trapped
  ions},}\ }\href
  {https://journals.aps.org/prl/abstract/10.1103/PhysRevLett.112.100403}
  {\bibfield  {journal} {\bibinfo  {journal} {Phys. Rev. Lett.}\ }\textbf
  {\bibinfo {volume} {112}},\ \bibinfo {pages} {100403} (\bibinfo {year}
  {2014})}\BibitemShut {NoStop}%
\bibitem [{\citenamefont {Walther}\ \emph {et~al.}(2005)\citenamefont
  {Walther}, \citenamefont {Aspelmeyer}, \citenamefont {Resch},\ and\
  \citenamefont {Zeilinger}}]{Walther2005}%
  \BibitemOpen
  \bibfield  {author} {\bibinfo {author} {\bibfnamefont {P.}~\bibnamefont
  {Walther}}, \bibinfo {author} {\bibfnamefont {M.}~\bibnamefont {Aspelmeyer}},
  \bibinfo {author} {\bibfnamefont {K.}~\bibnamefont {Resch}}, \ and\ \bibinfo
  {author} {\bibfnamefont {A.}~\bibnamefont {Zeilinger}},\ }\bibfield  {title}
  {\enquote {\bibinfo {title} {Experimental violation of a cluster state {Bell}
  inequality},}\ }\href
  {https://journals.aps.org/prl/abstract/10.1103/PhysRevLett.95.020403}
  {\bibfield  {journal} {\bibinfo  {journal} {Phys. Rev. Lett.}\ }\textbf
  {\bibinfo {volume} {95}},\ \bibinfo {pages} {020403} (\bibinfo {year}
  {2005})}\BibitemShut {NoStop}%
\bibitem [{\citenamefont {Gao}\ \emph {et~al.}(2010)\citenamefont {Gao},
  \citenamefont {Yao}, \citenamefont {Xu}, \citenamefont {Lu}, \citenamefont
  {G\"uhne}, \citenamefont {Cabello}, \citenamefont {Lu}, \citenamefont {Yang},
  \citenamefont {Chen},\ and\ \citenamefont {Pan}}]{WBGao2010}%
  \BibitemOpen
  \bibfield  {author} {\bibinfo {author} {\bibfnamefont {W.-B.}\ \bibnamefont
  {Gao}}, \bibinfo {author} {\bibfnamefont {X.-C.}\ \bibnamefont {Yao}},
  \bibinfo {author} {\bibfnamefont {P.}~\bibnamefont {Xu}}, \bibinfo {author}
  {\bibfnamefont {H.}~\bibnamefont {Lu}}, \bibinfo {author} {\bibfnamefont
  {O.}~\bibnamefont {G\"uhne}}, \bibinfo {author} {\bibfnamefont
  {A.}~\bibnamefont {Cabello}}, \bibinfo {author} {\bibfnamefont {C.-Y.}\
  \bibnamefont {Lu}}, \bibinfo {author} {\bibfnamefont {T.}~\bibnamefont
  {Yang}}, \bibinfo {author} {\bibfnamefont {Z.-B.}\ \bibnamefont {Chen}}, \
  and\ \bibinfo {author} {\bibfnamefont {J.-W.}\ \bibnamefont {Pan}},\
  }\bibfield  {title} {\enquote {\bibinfo {title} {Bell inequality tests of
  four-photon six-qubit graph states},}\ }\href
  {https://journals.aps.org/pra/abstract/10.1103/PhysRevA.82.042334} {\bibfield
   {journal} {\bibinfo  {journal} {Phys. Rev. A}\ }\textbf {\bibinfo {volume}
  {82}},\ \bibinfo {pages} {042334} (\bibinfo {year} {2010})}\BibitemShut
  {NoStop}%
\bibitem [{\citenamefont {You}\ and\ \citenamefont {Nori}(2003)}]{You2003}%
  \BibitemOpen
  \bibfield  {author} {\bibinfo {author} {\bibfnamefont {J.~Q.}\ \bibnamefont
  {You}}\ and\ \bibinfo {author} {\bibfnamefont {F.}~\bibnamefont {Nori}},\
  }\bibfield  {title} {\enquote {\bibinfo {title} {Quantum information
  processing with superconducting qubits in a microwave field},}\ }\href
  {\doibase 10.1103/PhysRevB.68.064509} {\bibfield  {journal} {\bibinfo
  {journal} {Phys. Rev. B}\ }\textbf {\bibinfo {volume} {68}},\ \bibinfo
  {pages} {064509} (\bibinfo {year} {2003})}\BibitemShut {NoStop}%
\bibitem [{\citenamefont {Blais}\ \emph {et~al.}(2004)\citenamefont {Blais},
  \citenamefont {Huang}, \citenamefont {Wallraff}, \citenamefont {Girvin},\
  and\ \citenamefont {Schoelkopf}}]{Blais2004}%
  \BibitemOpen
  \bibfield  {author} {\bibinfo {author} {\bibfnamefont {A.}~\bibnamefont
  {Blais}}, \bibinfo {author} {\bibfnamefont {R.-S.}\ \bibnamefont {Huang}},
  \bibinfo {author} {\bibfnamefont {A.}~\bibnamefont {Wallraff}}, \bibinfo
  {author} {\bibfnamefont {S.~M.}\ \bibnamefont {Girvin}}, \ and\ \bibinfo
  {author} {\bibfnamefont {R.~J.}\ \bibnamefont {Schoelkopf}},\ }\bibfield
  {title} {\enquote {\bibinfo {title} {Cavity quantum electrodynamics for
  superconducting electrical circuits: An architecture for quantum
  computation},}\ }\href {\doibase 10.1103/PhysRevA.69.062320} {\bibfield
  {journal} {\bibinfo  {journal} {Phys. Rev. A}\ }\textbf {\bibinfo {volume}
  {69}},\ \bibinfo {pages} {062320} (\bibinfo {year} {2004})}\BibitemShut
  {NoStop}%
\bibitem [{\citenamefont {Wallraff}\ \emph {et~al.}(2004)\citenamefont
  {Wallraff}, \citenamefont {Schuster}, \citenamefont {Blais}, \citenamefont
  {Frunzio}, \citenamefont {Huang}, \citenamefont {Majer}, \citenamefont
  {Kumar}, \citenamefont {Girvin},\ and\ \citenamefont
  {Schoelkopf}}]{Wallraff2004}%
  \BibitemOpen
  \bibfield  {author} {\bibinfo {author} {\bibfnamefont {A.}~\bibnamefont
  {Wallraff}}, \bibinfo {author} {\bibfnamefont {D.~I.}\ \bibnamefont
  {Schuster}}, \bibinfo {author} {\bibfnamefont {A.}~\bibnamefont {Blais}},
  \bibinfo {author} {\bibfnamefont {L.}~\bibnamefont {Frunzio}}, \bibinfo
  {author} {\bibfnamefont {R.-S.}\ \bibnamefont {Huang}}, \bibinfo {author}
  {\bibfnamefont {J.}~\bibnamefont {Majer}}, \bibinfo {author} {\bibfnamefont
  {S.}~\bibnamefont {Kumar}}, \bibinfo {author} {\bibfnamefont {S.~M.}\
  \bibnamefont {Girvin}}, \ and\ \bibinfo {author} {\bibfnamefont {R.~J.}\
  \bibnamefont {Schoelkopf}},\ }\bibfield  {title} {\enquote {\bibinfo {title}
  {Strong coupling of a single photon to a superconducting qubit using circuit
  quantum electrodynamics},}\ }\href
  {https://www.nature.com/articles/nature02851} {\bibfield  {journal} {\bibinfo
   {journal} {Nature}\ }\textbf {\bibinfo {volume} {431}},\ \bibinfo {pages}
  {162} (\bibinfo {year} {2004})}\BibitemShut {NoStop}%
\bibitem [{\citenamefont {You}\ and\ \citenamefont {Nori}(2011)}]{You2011}%
  \BibitemOpen
  \bibfield  {author} {\bibinfo {author} {\bibfnamefont {J.~Q.}\ \bibnamefont
  {You}}\ and\ \bibinfo {author} {\bibfnamefont {F.}~\bibnamefont {Nori}},\
  }\bibfield  {title} {\enquote {\bibinfo {title} {Atomic physics and quantum
  optics using superconducting circuits},}\ }\href
  {https://www.nature.com/articles/nature10122} {\bibfield  {journal} {\bibinfo
   {journal} {Nature}\ }\textbf {\bibinfo {volume} {474}},\ \bibinfo {pages}
  {589} (\bibinfo {year} {2011})}\BibitemShut {NoStop}%
\bibitem [{\citenamefont {Devoret}\ and\ \citenamefont
  {Schoelkopf}(2013)}]{Devoret2013}%
  \BibitemOpen
  \bibfield  {author} {\bibinfo {author} {\bibfnamefont {M.~H.}\ \bibnamefont
  {Devoret}}\ and\ \bibinfo {author} {\bibfnamefont {R.~J.}\ \bibnamefont
  {Schoelkopf}},\ }\bibfield  {title} {\enquote {\bibinfo {title}
  {Superconducting circuits for quantum information: An outlook},}\ }\href
  {https://science.sciencemag.org/content/339/6124/1169} {\bibfield  {journal}
  {\bibinfo  {journal} {Science}\ }\textbf {\bibinfo {volume} {339}},\ \bibinfo
  {pages} {1169} (\bibinfo {year} {2013})}\BibitemShut {NoStop}%
\bibitem [{\citenamefont {Gu}\ \emph {et~al.}(2017)\citenamefont {Gu},
  \citenamefont {Kockum}, \citenamefont {Miranowicz}, \citenamefont {Liu},\
  and\ \citenamefont {Nori}}]{Gu2017Microwave}%
  \BibitemOpen
  \bibfield  {author} {\bibinfo {author} {\bibfnamefont {X.}~\bibnamefont
  {Gu}}, \bibinfo {author} {\bibfnamefont {A.~F.}\ \bibnamefont {Kockum}},
  \bibinfo {author} {\bibfnamefont {A.}~\bibnamefont {Miranowicz}}, \bibinfo
  {author} {\bibfnamefont {Y.~X.}\ \bibnamefont {Liu}}, \ and\ \bibinfo
  {author} {\bibfnamefont {F.}~\bibnamefont {Nori}},\ }\bibfield  {title}
  {\enquote {\bibinfo {title} {Microwave photonics with superconducting quantum
  circuits},}\ }\href {https://doi.org/10.1016/j.physrep.2017.10.002}
  {\bibfield  {journal} {\bibinfo  {journal} {Phys. Rep.}\ }\textbf {\bibinfo
  {volume} {718-719}},\ \bibinfo {pages} {1} (\bibinfo {year}
  {2017})}\BibitemShut {NoStop}%
\bibitem [{\citenamefont {Wang}\ \emph {et~al.}(2011)\citenamefont {Wang},
  \citenamefont {Mariantoni}, \citenamefont {Bialczak}, \citenamefont
  {Lenander}, \citenamefont {Lucero}, \citenamefont {Neeley}, \citenamefont
  {O'Connell}, \citenamefont {Sank}, \citenamefont {Weides}, \citenamefont
  {Wenner}, \citenamefont {Yamamoto}, \citenamefont {Yin}, \citenamefont
  {Zhao}, \citenamefont {Martinis},\ and\ \citenamefont {Cleland}}]{HWang2011}%
  \BibitemOpen
  \bibfield  {author} {\bibinfo {author} {\bibfnamefont {H.}~\bibnamefont
  {Wang}}, \bibinfo {author} {\bibfnamefont {M.}~\bibnamefont {Mariantoni}},
  \bibinfo {author} {\bibfnamefont {R.~C.}\ \bibnamefont {Bialczak}}, \bibinfo
  {author} {\bibfnamefont {M.}~\bibnamefont {Lenander}}, \bibinfo {author}
  {\bibfnamefont {E.}~\bibnamefont {Lucero}}, \bibinfo {author} {\bibfnamefont
  {M.}~\bibnamefont {Neeley}}, \bibinfo {author} {\bibfnamefont {A.~D.}\
  \bibnamefont {O'Connell}}, \bibinfo {author} {\bibfnamefont {D.}~\bibnamefont
  {Sank}}, \bibinfo {author} {\bibfnamefont {M.}~\bibnamefont {Weides}},
  \bibinfo {author} {\bibfnamefont {J.}~\bibnamefont {Wenner}}, \bibinfo
  {author} {\bibfnamefont {T.}~\bibnamefont {Yamamoto}}, \bibinfo {author}
  {\bibfnamefont {Y.}~\bibnamefont {Yin}}, \bibinfo {author} {\bibfnamefont
  {J.}~\bibnamefont {Zhao}}, \bibinfo {author} {\bibfnamefont {J.~M.}\
  \bibnamefont {Martinis}}, \ and\ \bibinfo {author} {\bibfnamefont {A.~N.}\
  \bibnamefont {Cleland}},\ }\bibfield  {title} {\enquote {\bibinfo {title}
  {Deterministic entanglement of photons in two superconducting microwave
  resonators},}\ }\href
  {https://journals.aps.org/prl/abstract/10.1103/PhysRevLett.106.060401}
  {\bibfield  {journal} {\bibinfo  {journal} {Phys. Rev. Lett.}\ }\textbf
  {\bibinfo {volume} {106}},\ \bibinfo {pages} {060401} (\bibinfo {year}
  {2011})}\BibitemShut {NoStop}%
\bibitem [{\citenamefont {Gao}\ \emph {et~al.}(2019)\citenamefont {Gao},
  \citenamefont {Lester}, \citenamefont {Chou}, \citenamefont {Frunzio},
  \citenamefont {Devoret}, \citenamefont {Jiang}, \citenamefont {Girvin},\ and\
  \citenamefont {Schoelkopf}}]{YYGao}%
  \BibitemOpen
  \bibfield  {author} {\bibinfo {author} {\bibfnamefont {Y.~Y.}\ \bibnamefont
  {Gao}}, \bibinfo {author} {\bibfnamefont {B.~J.}\ \bibnamefont {Lester}},
  \bibinfo {author} {\bibfnamefont {K.~S.}\ \bibnamefont {Chou}}, \bibinfo
  {author} {\bibfnamefont {L.}~\bibnamefont {Frunzio}}, \bibinfo {author}
  {\bibfnamefont {M.~H.}\ \bibnamefont {Devoret}}, \bibinfo {author}
  {\bibfnamefont {L.}~\bibnamefont {Jiang}}, \bibinfo {author} {\bibfnamefont
  {S.~M.}\ \bibnamefont {Girvin}}, \ and\ \bibinfo {author} {\bibfnamefont
  {R.~J.}\ \bibnamefont {Schoelkopf}},\ }\bibfield  {title} {\enquote {\bibinfo
  {title} {Entanglement of bosonic modes through an engineered exchange
  interaction},}\ }\href {https://www.nature.com/articles/s41586-019-0970-4}
  {\bibfield  {journal} {\bibinfo  {journal} {Nature}\ }\textbf {\bibinfo
  {volume} {566}},\ \bibinfo {pages} {509} (\bibinfo {year}
  {2019})}\BibitemShut {NoStop}%
\bibitem [{\citenamefont {Wang}\ \emph
  {et~al.}(2016{\natexlab{b}})\citenamefont {Wang}, \citenamefont {Gao},
  \citenamefont {Reinhold}, \citenamefont {Heeres}, \citenamefont {Ofek},
  \citenamefont {Chou}, \citenamefont {Axline}, \citenamefont {Reagor},
  \citenamefont {Blumoff}, \citenamefont {Sliwa}, \citenamefont {Frunzio},
  \citenamefont {Girvin}, \citenamefont {Jiang}, \citenamefont {Mirrahimi},
  \citenamefont {Devoret},\ and\ \citenamefont {Schoelkopf}}]{CWang2016}%
  \BibitemOpen
  \bibfield  {author} {\bibinfo {author} {\bibfnamefont {C.}~\bibnamefont
  {Wang}}, \bibinfo {author} {\bibfnamefont {Y.~Y.}\ \bibnamefont {Gao}},
  \bibinfo {author} {\bibfnamefont {P.}~\bibnamefont {Reinhold}}, \bibinfo
  {author} {\bibfnamefont {R.~W.}\ \bibnamefont {Heeres}}, \bibinfo {author}
  {\bibfnamefont {N.}~\bibnamefont {Ofek}}, \bibinfo {author} {\bibfnamefont
  {K.}~\bibnamefont {Chou}}, \bibinfo {author} {\bibfnamefont {C.}~\bibnamefont
  {Axline}}, \bibinfo {author} {\bibfnamefont {M.}~\bibnamefont {Reagor}},
  \bibinfo {author} {\bibfnamefont {J.}~\bibnamefont {Blumoff}}, \bibinfo
  {author} {\bibfnamefont {K.~M.}\ \bibnamefont {Sliwa}}, \bibinfo {author}
  {\bibfnamefont {L.}~\bibnamefont {Frunzio}}, \bibinfo {author} {\bibfnamefont
  {S.~M.}\ \bibnamefont {Girvin}}, \bibinfo {author} {\bibfnamefont
  {L.}~\bibnamefont {Jiang}}, \bibinfo {author} {\bibfnamefont
  {M.}~\bibnamefont {Mirrahimi}}, \bibinfo {author} {\bibfnamefont {M.~H.}\
  \bibnamefont {Devoret}}, \ and\ \bibinfo {author} {\bibfnamefont {R.~J.}\
  \bibnamefont {Schoelkopf}},\ }\bibfield  {title} {\enquote {\bibinfo {title}
  {A {Schr\"odinger} cat living in two boxes},}\ }\href
  {https://science.sciencemag.org/content/352/6289/1087} {\bibfield  {journal}
  {\bibinfo  {journal} {Science}\ }\textbf {\bibinfo {volume} {352}},\ \bibinfo
  {pages} {1087} (\bibinfo {year} {2016}{\natexlab{b}})}\BibitemShut {NoStop}%
\bibitem [{\citenamefont {Paik}\ \emph {et~al.}(2016)\citenamefont {Paik},
  \citenamefont {Mezzacapo}, \citenamefont {Sandberg}, \citenamefont {McClure},
  \citenamefont {Abdo}, \citenamefont {C\'orcoles}, \citenamefont {Dial},
  \citenamefont {Bogorin}, \citenamefont {Plourde}, \citenamefont {Steffen},
  \citenamefont {Cross}, \citenamefont {Gambetta},\ and\ \citenamefont
  {Chow}}]{Paik2016}%
  \BibitemOpen
  \bibfield  {author} {\bibinfo {author} {\bibfnamefont {H.}~\bibnamefont
  {Paik}}, \bibinfo {author} {\bibfnamefont {A.}~\bibnamefont {Mezzacapo}},
  \bibinfo {author} {\bibfnamefont {M.}~\bibnamefont {Sandberg}}, \bibinfo
  {author} {\bibfnamefont {D.}~\bibnamefont {McClure}}, \bibinfo {author}
  {\bibfnamefont {B.}~\bibnamefont {Abdo}}, \bibinfo {author} {\bibfnamefont
  {A.}~\bibnamefont {C\'orcoles}}, \bibinfo {author} {\bibfnamefont
  {O.}~\bibnamefont {Dial}}, \bibinfo {author} {\bibfnamefont {D.}~\bibnamefont
  {Bogorin}}, \bibinfo {author} {\bibfnamefont {B.}~\bibnamefont {Plourde}},
  \bibinfo {author} {\bibfnamefont {M.}~\bibnamefont {Steffen}}, \bibinfo
  {author} {\bibfnamefont {A.}~\bibnamefont {Cross}}, \bibinfo {author}
  {\bibfnamefont {J.}~\bibnamefont {Gambetta}}, \ and\ \bibinfo {author}
  {\bibfnamefont {J.~M.}\ \bibnamefont {Chow}},\ }\bibfield  {title} {\enquote
  {\bibinfo {title} {Experimental demonstration of a resonator-induced phase
  gate in a multiqubit circuit-{QED} system},}\ }\href
  {https://journals.aps.org/prl/abstract/10.1103/PhysRevLett.117.250502}
  {\bibfield  {journal} {\bibinfo  {journal} {Phys. Rev. Lett.}\ }\textbf
  {\bibinfo {volume} {117}} (\bibinfo {year} {2016})}\BibitemShut {NoStop}%
\bibitem [{\citenamefont {Song}\ \emph {et~al.}(2017)\citenamefont {Song},
  \citenamefont {Xu}, \citenamefont {Liu}, \citenamefont {ping Yang},
  \citenamefont {Zheng}, \citenamefont {Deng}, \citenamefont {Xie},
  \citenamefont {Huang}, \citenamefont {Guo}, \citenamefont {Zhang},
  \citenamefont {Zhang}, \citenamefont {Xu}, \citenamefont {Zheng},
  \citenamefont {Zhu}, \citenamefont {Wang}, \citenamefont {Chen},
  \citenamefont {Lu}, \citenamefont {Han},\ and\ \citenamefont
  {Pan}}]{CSong2017}%
  \BibitemOpen
  \bibfield  {author} {\bibinfo {author} {\bibfnamefont {C.}~\bibnamefont
  {Song}}, \bibinfo {author} {\bibfnamefont {K.}~\bibnamefont {Xu}}, \bibinfo
  {author} {\bibfnamefont {W.}~\bibnamefont {Liu}}, \bibinfo {author}
  {\bibfnamefont {C.}~\bibnamefont {ping Yang}}, \bibinfo {author}
  {\bibfnamefont {S.-B.}\ \bibnamefont {Zheng}}, \bibinfo {author}
  {\bibfnamefont {H.}~\bibnamefont {Deng}}, \bibinfo {author} {\bibfnamefont
  {Q.}~\bibnamefont {Xie}}, \bibinfo {author} {\bibfnamefont {K.}~\bibnamefont
  {Huang}}, \bibinfo {author} {\bibfnamefont {Q.}~\bibnamefont {Guo}}, \bibinfo
  {author} {\bibfnamefont {L.}~\bibnamefont {Zhang}}, \bibinfo {author}
  {\bibfnamefont {P.}~\bibnamefont {Zhang}}, \bibinfo {author} {\bibfnamefont
  {D.}~\bibnamefont {Xu}}, \bibinfo {author} {\bibfnamefont {D.}~\bibnamefont
  {Zheng}}, \bibinfo {author} {\bibfnamefont {X.}~\bibnamefont {Zhu}}, \bibinfo
  {author} {\bibfnamefont {H.}~\bibnamefont {Wang}}, \bibinfo {author}
  {\bibfnamefont {Y.-A.}\ \bibnamefont {Chen}}, \bibinfo {author}
  {\bibfnamefont {C.-Y.}\ \bibnamefont {Lu}}, \bibinfo {author} {\bibfnamefont
  {S.}~\bibnamefont {Han}}, \ and\ \bibinfo {author} {\bibfnamefont {J.-W.}\
  \bibnamefont {Pan}},\ }\bibfield  {title} {\enquote {\bibinfo {title}
  {10-qubit entanglement and parallel logic operations with a superconducting
  circuit},}\ }\href
  {https://journals.aps.org/prl/abstract/10.1103/PhysRevLett.119.180511}
  {\bibfield  {journal} {\bibinfo  {journal} {Phys. Rev. Lett.}\ }\textbf
  {\bibinfo {volume} {119}} (\bibinfo {year} {2017})}\BibitemShut {NoStop}%
\bibitem [{\citenamefont {Song}\ \emph {et~al.}(2019)\citenamefont {Song},
  \citenamefont {Xu}, \citenamefont {Li}, \citenamefont {Zhang}, \citenamefont
  {Zhang}, \citenamefont {Liu}, \citenamefont {Guo}, \citenamefont {Wang},
  \citenamefont {Ren}, \citenamefont {Hao}, \citenamefont {Feng}, \citenamefont
  {Fan}, \citenamefont {Zheng}, \citenamefont {Wang}, \citenamefont {Wang},\
  and\ \citenamefont {Zhu}}]{Song2019}%
  \BibitemOpen
  \bibfield  {author} {\bibinfo {author} {\bibfnamefont {C.}~\bibnamefont
  {Song}}, \bibinfo {author} {\bibfnamefont {K.}~\bibnamefont {Xu}}, \bibinfo
  {author} {\bibfnamefont {H.}~\bibnamefont {Li}}, \bibinfo {author}
  {\bibfnamefont {Y.-R.}\ \bibnamefont {Zhang}}, \bibinfo {author}
  {\bibfnamefont {X.}~\bibnamefont {Zhang}}, \bibinfo {author} {\bibfnamefont
  {W.}~\bibnamefont {Liu}}, \bibinfo {author} {\bibfnamefont {Q.}~\bibnamefont
  {Guo}}, \bibinfo {author} {\bibfnamefont {Z.}~\bibnamefont {Wang}}, \bibinfo
  {author} {\bibfnamefont {W.}~\bibnamefont {Ren}}, \bibinfo {author}
  {\bibfnamefont {J.}~\bibnamefont {Hao}}, \bibinfo {author} {\bibfnamefont
  {H.}~\bibnamefont {Feng}}, \bibinfo {author} {\bibfnamefont {H.}~\bibnamefont
  {Fan}}, \bibinfo {author} {\bibfnamefont {D.}~\bibnamefont {Zheng}}, \bibinfo
  {author} {\bibfnamefont {D.-W.}\ \bibnamefont {Wang}}, \bibinfo {author}
  {\bibfnamefont {H.}~\bibnamefont {Wang}}, \ and\ \bibinfo {author}
  {\bibfnamefont {S.-Y.}\ \bibnamefont {Zhu}},\ }\bibfield  {title} {\enquote
  {\bibinfo {title} {Generation of multicomponent atomic schrödinger cat states
  of up to 20 qubits},}\ }\href
  {https://science.sciencemag.org/content/365/6453/574.editor-summary}
  {\bibfield  {journal} {\bibinfo  {journal} {Science}\ }\textbf {\bibinfo
  {volume} {365}},\ \bibinfo {pages} {574} (\bibinfo {year}
  {2019})}\BibitemShut {NoStop}%
\bibitem [{\citenamefont {Ansmann}\ \emph {et~al.}(2009)\citenamefont
  {Ansmann}, \citenamefont {Wang}, \citenamefont {Bialczak}, \citenamefont
  {Hofheinz}, \citenamefont {Lucero}, \citenamefont {Neeley}, \citenamefont
  {O'Connell}, \citenamefont {Sank}, \citenamefont {Weides}, \citenamefont
  {Wenner}, \citenamefont {Cleland},\ and\ \citenamefont
  {Martinis}}]{Ansmann2009}%
  \BibitemOpen
  \bibfield  {author} {\bibinfo {author} {\bibfnamefont {M.}~\bibnamefont
  {Ansmann}}, \bibinfo {author} {\bibfnamefont {H.}~\bibnamefont {Wang}},
  \bibinfo {author} {\bibfnamefont {R.~C.}\ \bibnamefont {Bialczak}}, \bibinfo
  {author} {\bibfnamefont {M.}~\bibnamefont {Hofheinz}}, \bibinfo {author}
  {\bibfnamefont {E.}~\bibnamefont {Lucero}}, \bibinfo {author} {\bibfnamefont
  {M.}~\bibnamefont {Neeley}}, \bibinfo {author} {\bibfnamefont {A.~D.}\
  \bibnamefont {O'Connell}}, \bibinfo {author} {\bibfnamefont {D.}~\bibnamefont
  {Sank}}, \bibinfo {author} {\bibfnamefont {M.}~\bibnamefont {Weides}},
  \bibinfo {author} {\bibfnamefont {J.}~\bibnamefont {Wenner}}, \bibinfo
  {author} {\bibfnamefont {A.~N.}\ \bibnamefont {Cleland}}, \ and\ \bibinfo
  {author} {\bibfnamefont {J.~M.}\ \bibnamefont {Martinis}},\ }\bibfield
  {title} {\enquote {\bibinfo {title} {Violation of bell's inequality in
  josephson phase qubits},}\ }\href {\doibase 10.1038/nature08363} {\bibfield
  {journal} {\bibinfo  {journal} {Nature}\ }\textbf {\bibinfo {volume} {461}},\
  \bibinfo {pages} {504} (\bibinfo {year} {2009})}\BibitemShut {NoStop}%
\bibitem [{\citenamefont {Chow}\ \emph {et~al.}(2010)\citenamefont {Chow},
  \citenamefont {DiCarlo}, \citenamefont {Gambetta}, \citenamefont
  {Nunnenkamp}, \citenamefont {Bishop}, \citenamefont {Frunzio}, \citenamefont
  {Devoret}, \citenamefont {Girvin},\ and\ \citenamefont
  {Schoelkopf}}]{ChowPRA2010}%
  \BibitemOpen
  \bibfield  {author} {\bibinfo {author} {\bibfnamefont {J.~M.}\ \bibnamefont
  {Chow}}, \bibinfo {author} {\bibfnamefont {L.}~\bibnamefont {DiCarlo}},
  \bibinfo {author} {\bibfnamefont {J.~M.}\ \bibnamefont {Gambetta}}, \bibinfo
  {author} {\bibfnamefont {A.}~\bibnamefont {Nunnenkamp}}, \bibinfo {author}
  {\bibfnamefont {L.~S.}\ \bibnamefont {Bishop}}, \bibinfo {author}
  {\bibfnamefont {L.}~\bibnamefont {Frunzio}}, \bibinfo {author} {\bibfnamefont
  {M.~H.}\ \bibnamefont {Devoret}}, \bibinfo {author} {\bibfnamefont {S.~M.}\
  \bibnamefont {Girvin}}, \ and\ \bibinfo {author} {\bibfnamefont {R.~J.}\
  \bibnamefont {Schoelkopf}},\ }\bibfield  {title} {\enquote {\bibinfo {title}
  {Detecting highly entangled states with a joint qubit readout},}\ }\href
  {\doibase 10.1103/PhysRevA.81.062325} {\bibfield  {journal} {\bibinfo
  {journal} {Phys. Rev. A}\ }\textbf {\bibinfo {volume} {81}},\ \bibinfo
  {pages} {062325} (\bibinfo {year} {2010})}\BibitemShut {NoStop}%
\bibitem [{\citenamefont {Zhong}\ \emph {et~al.}(2019)\citenamefont {Zhong},
  \citenamefont {Chang}, \citenamefont {Satzinger}, \citenamefont {Chou},
  \citenamefont {Bienfait}, \citenamefont {Conner}, \citenamefont {Dumur},
  \citenamefont {Grebel}, \citenamefont {Peairs}, \citenamefont {Povey},
  \citenamefont {Schuster},\ and\ \citenamefont {Cleland}}]{Zhong2019}%
  \BibitemOpen
  \bibfield  {author} {\bibinfo {author} {\bibfnamefont {Y.~P.}\ \bibnamefont
  {Zhong}}, \bibinfo {author} {\bibfnamefont {H.-S.}\ \bibnamefont {Chang}},
  \bibinfo {author} {\bibfnamefont {K.~J.}\ \bibnamefont {Satzinger}}, \bibinfo
  {author} {\bibfnamefont {M.-H.}\ \bibnamefont {Chou}}, \bibinfo {author}
  {\bibfnamefont {A.}~\bibnamefont {Bienfait}}, \bibinfo {author}
  {\bibfnamefont {C.~R.}\ \bibnamefont {Conner}}, \bibinfo {author}
  {\bibfnamefont {E.}~\bibnamefont {Dumur}}, \bibinfo {author} {\bibfnamefont
  {J.}~\bibnamefont {Grebel}}, \bibinfo {author} {\bibfnamefont {G.~A.}\
  \bibnamefont {Peairs}}, \bibinfo {author} {\bibfnamefont {R.~G.}\
  \bibnamefont {Povey}}, \bibinfo {author} {\bibfnamefont {D.~I.}\ \bibnamefont
  {Schuster}}, \ and\ \bibinfo {author} {\bibfnamefont {A.~N.}\ \bibnamefont
  {Cleland}},\ }\bibfield  {title} {\enquote {\bibinfo {title} {Violating
  {B}ell's inequality with remotely connected superconducting qubits},}\ }\href
  {https://www.nature.com/articles/s41567-019-0507-7} {\bibfield  {journal}
  {\bibinfo  {journal} {Nat. Phys.}\ }\textbf {\bibinfo {volume} {15}},\
  \bibinfo {pages} {741} (\bibinfo {year} {2019})}\BibitemShut {NoStop}%
\bibitem [{\citenamefont {Vlastakis}\ \emph {et~al.}(2015)\citenamefont
  {Vlastakis}, \citenamefont {Petrenko}, \citenamefont {Ofek}, \citenamefont
  {Sun}, \citenamefont {Leghtas}, \citenamefont {Sliwa}, \citenamefont {Liu},
  \citenamefont {Hatridge}, \citenamefont {Blumoff}, \citenamefont {Frunzio},
  \citenamefont {Mirrahimi}, \citenamefont {Jiang}, \citenamefont {Devoret},\
  and\ \citenamefont {Schoelkopf}}]{Vlastakis2015}%
  \BibitemOpen
  \bibfield  {author} {\bibinfo {author} {\bibfnamefont {B.}~\bibnamefont
  {Vlastakis}}, \bibinfo {author} {\bibfnamefont {A.}~\bibnamefont {Petrenko}},
  \bibinfo {author} {\bibfnamefont {N.}~\bibnamefont {Ofek}}, \bibinfo {author}
  {\bibfnamefont {L.}~\bibnamefont {Sun}}, \bibinfo {author} {\bibfnamefont
  {Z.}~\bibnamefont {Leghtas}}, \bibinfo {author} {\bibfnamefont
  {K.}~\bibnamefont {Sliwa}}, \bibinfo {author} {\bibfnamefont
  {Y.}~\bibnamefont {Liu}}, \bibinfo {author} {\bibfnamefont {M.}~\bibnamefont
  {Hatridge}}, \bibinfo {author} {\bibfnamefont {J.}~\bibnamefont {Blumoff}},
  \bibinfo {author} {\bibfnamefont {L.}~\bibnamefont {Frunzio}}, \bibinfo
  {author} {\bibfnamefont {M.}~\bibnamefont {Mirrahimi}}, \bibinfo {author}
  {\bibfnamefont {L.}~\bibnamefont {Jiang}}, \bibinfo {author} {\bibfnamefont
  {M.~H.}\ \bibnamefont {Devoret}}, \ and\ \bibinfo {author} {\bibfnamefont
  {R.~J.}\ \bibnamefont {Schoelkopf}},\ }\bibfield  {title} {\enquote {\bibinfo
  {title} {Characterizing entanglement of an artificial atom and a cavity cat
  state with {Bell's} inequality},}\ }\href
  {https://www.nature.com/articles/ncomms9970} {\bibfield  {journal} {\bibinfo
  {journal} {Nat. Commun.}\ }\textbf {\bibinfo {volume} {6}},\ \bibinfo {pages}
  {8970} (\bibinfo {year} {2015})}\BibitemShut {NoStop}%
\bibitem [{\citenamefont {Paik}\ \emph {et~al.}(2011)\citenamefont {Paik},
  \citenamefont {Schuster}, \citenamefont {Bishop}, \citenamefont {Kirchmair},
  \citenamefont {Catelani}, \citenamefont {Sears}, \citenamefont {Johnson},
  \citenamefont {Reagor}, \citenamefont {Frunzio}, \citenamefont {Glazman},
  \citenamefont {Girvin}, \citenamefont {Devoret},\ and\ \citenamefont
  {Schoelkopf}}]{Paik2011}%
  \BibitemOpen
  \bibfield  {author} {\bibinfo {author} {\bibfnamefont {H.}~\bibnamefont
  {Paik}}, \bibinfo {author} {\bibfnamefont {D.~I.}\ \bibnamefont {Schuster}},
  \bibinfo {author} {\bibfnamefont {L.~S.}\ \bibnamefont {Bishop}}, \bibinfo
  {author} {\bibfnamefont {G.}~\bibnamefont {Kirchmair}}, \bibinfo {author}
  {\bibfnamefont {G.}~\bibnamefont {Catelani}}, \bibinfo {author}
  {\bibfnamefont {A.~P.}\ \bibnamefont {Sears}}, \bibinfo {author}
  {\bibfnamefont {B.~R.}\ \bibnamefont {Johnson}}, \bibinfo {author}
  {\bibfnamefont {M.~J.}\ \bibnamefont {Reagor}}, \bibinfo {author}
  {\bibfnamefont {L.}~\bibnamefont {Frunzio}}, \bibinfo {author} {\bibfnamefont
  {L.~I.}\ \bibnamefont {Glazman}}, \bibinfo {author} {\bibfnamefont {S.~M.}\
  \bibnamefont {Girvin}}, \bibinfo {author} {\bibfnamefont {M.~H.}\
  \bibnamefont {Devoret}}, \ and\ \bibinfo {author} {\bibfnamefont {R.~J.}\
  \bibnamefont {Schoelkopf}},\ }\bibfield  {title} {\enquote {\bibinfo {title}
  {Observation of high coherence in {J}osephson junction qubits measured in a
  three-dimensional circuit qed architecture},}\ }\href {\doibase
  10.1103/PhysRevLett.107.240501} {\bibfield  {journal} {\bibinfo  {journal}
  {Phys. Rev. Lett.}\ }\textbf {\bibinfo {volume} {107}},\ \bibinfo {pages}
  {240501} (\bibinfo {year} {2011})}\BibitemShut {NoStop}%
\bibitem [{\citenamefont {Ma}\ \emph {et~al.}(2019)\citenamefont {Ma},
  \citenamefont {Xu}, \citenamefont {Mu}, \citenamefont {Cai}, \citenamefont
  {Hu}, \citenamefont {Wang}, \citenamefont {Pan}, \citenamefont {Wang},
  \citenamefont {Song}, \citenamefont {Zou},\ and\ \citenamefont
  {Sun}}]{Ma2019}%
  \BibitemOpen
  \bibfield  {author} {\bibinfo {author} {\bibfnamefont {Y.}~\bibnamefont
  {Ma}}, \bibinfo {author} {\bibfnamefont {Y.}~\bibnamefont {Xu}}, \bibinfo
  {author} {\bibfnamefont {X.}~\bibnamefont {Mu}}, \bibinfo {author}
  {\bibfnamefont {W.}~\bibnamefont {Cai}}, \bibinfo {author} {\bibfnamefont
  {L.}~\bibnamefont {Hu}}, \bibinfo {author} {\bibfnamefont {W.}~\bibnamefont
  {Wang}}, \bibinfo {author} {\bibfnamefont {X.}~\bibnamefont {Pan}}, \bibinfo
  {author} {\bibfnamefont {H.}~\bibnamefont {Wang}}, \bibinfo {author}
  {\bibfnamefont {Y.~P.}\ \bibnamefont {Song}}, \bibinfo {author}
  {\bibfnamefont {C.~L.}\ \bibnamefont {Zou}}, \ and\ \bibinfo {author}
  {\bibfnamefont {L.}~\bibnamefont {Sun}},\ }\bibfield  {title} {\enquote
  {\bibinfo {title} {Error-transparent operations on a logical qubit protected
  by quantum error correction},}\ }\href {https://arxiv.org/abs/1909.06803}
  {\bibfield  {journal} {\bibinfo  {journal} {arXiv:1909.06803}\ } (\bibinfo
  {year} {2019})}\BibitemShut {NoStop}%
\bibitem [{\citenamefont {Reinhold}\ \emph {et~al.}(2019)\citenamefont
  {Reinhold}, \citenamefont {Rosenblum}, \citenamefont {Ma}, \citenamefont
  {Frunzio}, \citenamefont {Jiang},\ and\ \citenamefont
  {Schoelkopf}}]{Reinhold2019}%
  \BibitemOpen
  \bibfield  {author} {\bibinfo {author} {\bibfnamefont {P.}~\bibnamefont
  {Reinhold}}, \bibinfo {author} {\bibfnamefont {S.}~\bibnamefont {Rosenblum}},
  \bibinfo {author} {\bibfnamefont {W.-L.}\ \bibnamefont {Ma}}, \bibinfo
  {author} {\bibfnamefont {L.}~\bibnamefont {Frunzio}}, \bibinfo {author}
  {\bibfnamefont {L.}~\bibnamefont {Jiang}}, \ and\ \bibinfo {author}
  {\bibfnamefont {R.~J.}\ \bibnamefont {Schoelkopf}},\ }\bibfield  {title}
  {\enquote {\bibinfo {title} {Error-corrected gates on an encoded qubit},}\
  }\href {https://arxiv.org/abs/1907.12327} {\bibfield  {journal} {\bibinfo
  {journal} {arXiv:1907.12327}\ } (\bibinfo {year} {2019})}\BibitemShut
  {NoStop}%
\bibitem [{\citenamefont {Xu}\ \emph {et~al.}(2020)\citenamefont {Xu},
  \citenamefont {Ma}, \citenamefont {Cai}, \citenamefont {Mu}, \citenamefont
  {Dai}, \citenamefont {Wang}, \citenamefont {Hu}, \citenamefont {Li},
  \citenamefont {Han}, \citenamefont {Wang}, \citenamefont {Song},
  \citenamefont {Yang}, \citenamefont {Zheng},\ and\ \citenamefont
  {Sun}}]{Xu2019}%
  \BibitemOpen
  \bibfield  {author} {\bibinfo {author} {\bibfnamefont {Y.}~\bibnamefont
  {Xu}}, \bibinfo {author} {\bibfnamefont {Y.}~\bibnamefont {Ma}}, \bibinfo
  {author} {\bibfnamefont {W.}~\bibnamefont {Cai}}, \bibinfo {author}
  {\bibfnamefont {X.}~\bibnamefont {Mu}}, \bibinfo {author} {\bibfnamefont
  {W.}~\bibnamefont {Dai}}, \bibinfo {author} {\bibfnamefont {W.}~\bibnamefont
  {Wang}}, \bibinfo {author} {\bibfnamefont {L.}~\bibnamefont {Hu}}, \bibinfo
  {author} {\bibfnamefont {X.}~\bibnamefont {Li}}, \bibinfo {author}
  {\bibfnamefont {J.}~\bibnamefont {Han}}, \bibinfo {author} {\bibfnamefont
  {H.}~\bibnamefont {Wang}}, \bibinfo {author} {\bibfnamefont {Y.~P.}\
  \bibnamefont {Song}}, \bibinfo {author} {\bibfnamefont {Z.-B.}\ \bibnamefont
  {Yang}}, \bibinfo {author} {\bibfnamefont {S.-B.}\ \bibnamefont {Zheng}}, \
  and\ \bibinfo {author} {\bibfnamefont {L.}~\bibnamefont {Sun}},\ }\bibfield
  {title} {\enquote {\bibinfo {title} {Demonstration of controlled-phase gates
  between two error-correctable photonic qubits},}\ }\href {\doibase
  10.1103/PhysRevLett.124.120501} {\bibfield  {journal} {\bibinfo  {journal}
  {Phys. Rev. Lett.}\ }\textbf {\bibinfo {volume} {124}},\ \bibinfo {pages}
  {120501} (\bibinfo {year} {2020})}\BibitemShut {NoStop}%
\bibitem [{Sup()}]{Supplement}%
  \BibitemOpen
  \href@noop {} {\bibinfo  {journal} {{Supplementary Materials}}\ }\BibitemShut
  {NoStop}%
\bibitem [{\citenamefont {Vlastakis}\ \emph {et~al.}(2013)\citenamefont
  {Vlastakis}, \citenamefont {Kirchmair}, \citenamefont {Leghtas},
  \citenamefont {Nigg}, \citenamefont {Frunzio}, \citenamefont {Girvin},
  \citenamefont {Mirrahimi}, \citenamefont {Devoret},\ and\ \citenamefont
  {Schoelkopf}}]{Vlastakis2013}%
  \BibitemOpen
\bibfield  {journal} {  }\bibfield  {author} {\bibinfo {author} {\bibfnamefont
  {B.}~\bibnamefont {Vlastakis}}, \bibinfo {author} {\bibfnamefont
  {G.}~\bibnamefont {Kirchmair}}, \bibinfo {author} {\bibfnamefont
  {Z.}~\bibnamefont {Leghtas}}, \bibinfo {author} {\bibfnamefont {S.~E.}\
  \bibnamefont {Nigg}}, \bibinfo {author} {\bibfnamefont {L.}~\bibnamefont
  {Frunzio}}, \bibinfo {author} {\bibfnamefont {S.~M.}\ \bibnamefont {Girvin}},
  \bibinfo {author} {\bibfnamefont {M.}~\bibnamefont {Mirrahimi}}, \bibinfo
  {author} {\bibfnamefont {M.~H.}\ \bibnamefont {Devoret}}, \ and\ \bibinfo
  {author} {\bibfnamefont {R.~J.}\ \bibnamefont {Schoelkopf}},\ }\bibfield
  {title} {\enquote {\bibinfo {title} {Deterministically encoding quantum
  information using 100-photon {Schr\"odinger} cat states},}\ }\href
  {https://science.sciencemag.org/content/342/6158/607} {\bibfield  {journal}
  {\bibinfo  {journal} {Science}\ }\textbf {\bibinfo {volume} {342}},\ \bibinfo
  {pages} {607} (\bibinfo {year} {2013})}\BibitemShut {NoStop}%
\bibitem [{\citenamefont {Sun}\ \emph {et~al.}(2014)\citenamefont {Sun},
  \citenamefont {Petrenko}, \citenamefont {Leghtas}, \citenamefont {Vlastakis},
  \citenamefont {Kirchmair}, \citenamefont {Sliwa}, \citenamefont {Narla},
  \citenamefont {Hatridge}, \citenamefont {Shankar}, \citenamefont {Blumoff},
  \citenamefont {Frunzio}, \citenamefont {Mirrahimi}, \citenamefont {Devoret},\
  and\ \citenamefont {Schoelkopf}}]{Sun2014}%
  \BibitemOpen
  \bibfield  {author} {\bibinfo {author} {\bibfnamefont {L.}~\bibnamefont
  {Sun}}, \bibinfo {author} {\bibfnamefont {A.}~\bibnamefont {Petrenko}},
  \bibinfo {author} {\bibfnamefont {Z.}~\bibnamefont {Leghtas}}, \bibinfo
  {author} {\bibfnamefont {B.}~\bibnamefont {Vlastakis}}, \bibinfo {author}
  {\bibfnamefont {G.}~\bibnamefont {Kirchmair}}, \bibinfo {author}
  {\bibfnamefont {K.~M.}\ \bibnamefont {Sliwa}}, \bibinfo {author}
  {\bibfnamefont {A.}~\bibnamefont {Narla}}, \bibinfo {author} {\bibfnamefont
  {M.}~\bibnamefont {Hatridge}}, \bibinfo {author} {\bibfnamefont
  {S.}~\bibnamefont {Shankar}}, \bibinfo {author} {\bibfnamefont
  {J.}~\bibnamefont {Blumoff}}, \bibinfo {author} {\bibfnamefont
  {L.}~\bibnamefont {Frunzio}}, \bibinfo {author} {\bibfnamefont
  {M.}~\bibnamefont {Mirrahimi}}, \bibinfo {author} {\bibfnamefont {M.~H.}\
  \bibnamefont {Devoret}}, \ and\ \bibinfo {author} {\bibfnamefont {R.~J.}\
  \bibnamefont {Schoelkopf}},\ }\bibfield  {title} {\enquote {\bibinfo {title}
  {Tracking photon jumps with repeated quantum non-demolition parity
  measurements},}\ }\href {https://www.nature.com/articles/nature13436}
  {\bibfield  {journal} {\bibinfo  {journal} {Nature}\ }\textbf {\bibinfo
  {volume} {511}},\ \bibinfo {pages} {444} (\bibinfo {year}
  {2014})}\BibitemShut {NoStop}%
\bibitem [{\citenamefont {Liu}\ \emph {et~al.}(2017)\citenamefont {Liu},
  \citenamefont {Xu}, \citenamefont {Wang}, \citenamefont {Zheng},
  \citenamefont {Roy}, \citenamefont {Kundu}, \citenamefont {Chand},
  \citenamefont {Ranadive}, \citenamefont {Vijay}, \citenamefont {Song} \emph
  {et~al.}}]{Liu2017}%
  \BibitemOpen
  \bibfield  {author} {\bibinfo {author} {\bibfnamefont {K.}~\bibnamefont
  {Liu}}, \bibinfo {author} {\bibfnamefont {Y.}~\bibnamefont {Xu}}, \bibinfo
  {author} {\bibfnamefont {W.}~\bibnamefont {Wang}}, \bibinfo {author}
  {\bibfnamefont {S.-B.}\ \bibnamefont {Zheng}}, \bibinfo {author}
  {\bibfnamefont {T.}~\bibnamefont {Roy}}, \bibinfo {author} {\bibfnamefont
  {S.}~\bibnamefont {Kundu}}, \bibinfo {author} {\bibfnamefont
  {M.}~\bibnamefont {Chand}}, \bibinfo {author} {\bibfnamefont
  {A.}~\bibnamefont {Ranadive}}, \bibinfo {author} {\bibfnamefont
  {R.}~\bibnamefont {Vijay}}, \bibinfo {author} {\bibfnamefont
  {Y.}~\bibnamefont {Song}},  \emph {et~al.},\ }\bibfield  {title} {\enquote
  {\bibinfo {title} {A twofold quantum delayed-choice experiment in a
  superconducting circuit},}\ }\href
  {http://advances.sciencemag.org/lookup/doi/10.1126/sciadv.1603159} {\bibfield
   {journal} {\bibinfo  {journal} {Sci. Adv.}\ }\textbf {\bibinfo {volume}
  {3}},\ \bibinfo {pages} {e1603159} (\bibinfo {year} {2017})}\BibitemShut
  {NoStop}%
\bibitem [{\citenamefont {Wang}\ \emph {et~al.}(2017)\citenamefont {Wang},
  \citenamefont {Hu}, \citenamefont {Xu}, \citenamefont {Liu}, \citenamefont
  {Ma}, \citenamefont {Zheng}, \citenamefont {Vijay}, \citenamefont {Song},
  \citenamefont {Duan},\ and\ \citenamefont {Sun}}]{WWang2017}%
  \BibitemOpen
  \bibfield  {author} {\bibinfo {author} {\bibfnamefont {W.}~\bibnamefont
  {Wang}}, \bibinfo {author} {\bibfnamefont {L.}~\bibnamefont {Hu}}, \bibinfo
  {author} {\bibfnamefont {Y.}~\bibnamefont {Xu}}, \bibinfo {author}
  {\bibfnamefont {K.}~\bibnamefont {Liu}}, \bibinfo {author} {\bibfnamefont
  {Y.}~\bibnamefont {Ma}}, \bibinfo {author} {\bibfnamefont {S.-B.}\
  \bibnamefont {Zheng}}, \bibinfo {author} {\bibfnamefont {R.}~\bibnamefont
  {Vijay}}, \bibinfo {author} {\bibfnamefont {Y.}~\bibnamefont {Song}},
  \bibinfo {author} {\bibfnamefont {L.-M.}\ \bibnamefont {Duan}}, \ and\
  \bibinfo {author} {\bibfnamefont {L.}~\bibnamefont {Sun}},\ }\bibfield
  {title} {\enquote {\bibinfo {title} {Converting quasiclassical states into
  arbitrary {Fock} state superpositions in a superconducting circuit},}\ }\href
  {https://journals.aps.org/prl/abstract/10.1103/PhysRevLett.118.223604}
  {\bibfield  {journal} {\bibinfo  {journal} {Phys. Rev. Lett.}\ }\textbf
  {\bibinfo {volume} {118}} (\bibinfo {year} {2017})}\BibitemShut {NoStop}%
\bibitem [{\citenamefont {Bertet}\ \emph {et~al.}(2002)\citenamefont {Bertet},
  \citenamefont {Auffeves}, \citenamefont {Maioli}, \citenamefont {Osnaghi},
  \citenamefont {Meunier}, \citenamefont {Brune}, \citenamefont {Raimond},\
  and\ \citenamefont {Haroche}}]{Bertet2002}%
  \BibitemOpen
  \bibfield  {author} {\bibinfo {author} {\bibfnamefont {P.}~\bibnamefont
  {Bertet}}, \bibinfo {author} {\bibfnamefont {A.}~\bibnamefont {Auffeves}},
  \bibinfo {author} {\bibfnamefont {P.}~\bibnamefont {Maioli}}, \bibinfo
  {author} {\bibfnamefont {S.}~\bibnamefont {Osnaghi}}, \bibinfo {author}
  {\bibfnamefont {T.}~\bibnamefont {Meunier}}, \bibinfo {author} {\bibfnamefont
  {M.}~\bibnamefont {Brune}}, \bibinfo {author} {\bibfnamefont {J.~M.}\
  \bibnamefont {Raimond}}, \ and\ \bibinfo {author} {\bibfnamefont
  {S.}~\bibnamefont {Haroche}},\ }\bibfield  {title} {\enquote {\bibinfo
  {title} {{Direct measurement of the Wigner function of a one-photon Fock
  state in a cavity}},}\ }\href
  {https://journals.aps.org/prl/abstract/10.1103/PhysRevLett.89.200402}
  {\bibfield  {journal} {\bibinfo  {journal} {Phys. Rev. Lett.}\ }\textbf
  {\bibinfo {volume} {89}} (\bibinfo {year} {2002})}\BibitemShut {NoStop}%
\bibitem [{\citenamefont {Milman}\ \emph {et~al.}(2005)\citenamefont {Milman},
  \citenamefont {Auffeves}, \citenamefont {Yamaguchi}, \citenamefont {M.Brune},
  \citenamefont {Raimond},\ and\ \citenamefont {Haroche}}]{Milman2005}%
  \BibitemOpen
  \bibfield  {author} {\bibinfo {author} {\bibfnamefont {P.}~\bibnamefont
  {Milman}}, \bibinfo {author} {\bibfnamefont {A.}~\bibnamefont {Auffeves}},
  \bibinfo {author} {\bibfnamefont {F.}~\bibnamefont {Yamaguchi}}, \bibinfo
  {author} {\bibnamefont {M.Brune}}, \bibinfo {author} {\bibfnamefont
  {J.}~\bibnamefont {Raimond}}, \ and\ \bibinfo {author} {\bibfnamefont
  {S.}~\bibnamefont {Haroche}},\ }\bibfield  {title} {\enquote {\bibinfo
  {title} {{A proposal to test Bell's inequalities with mesoscopic non-local
  states in cavity QED}},}\ }\href
  {https://link.springer.com/content/pdf/10.1140/epjd/e2004-00171-6} {\bibfield
   {journal} {\bibinfo  {journal} {The European Physical Journal D}\ }\textbf
  {\bibinfo {volume} {32}} (\bibinfo {year} {2005})}\BibitemShut {NoStop}%
\bibitem [{\citenamefont {Johansson}\ \emph {et~al.}(2012)\citenamefont
  {Johansson}, \citenamefont {Nation},\ and\ \citenamefont
  {Nori}}]{Johansson2012}%
  \BibitemOpen
  \bibfield  {author} {\bibinfo {author} {\bibfnamefont {J.~R.}\ \bibnamefont
  {Johansson}}, \bibinfo {author} {\bibfnamefont {P.~D.}\ \bibnamefont
  {Nation}}, \ and\ \bibinfo {author} {\bibfnamefont {F.}~\bibnamefont
  {Nori}},\ }\bibfield  {title} {\enquote {\bibinfo {title} {Qutip: An
  open-source python framework for the dynamics of open quantum systems},}\
  }\href {\doibase 10.1016/j.cpc.2012.02.021} {\bibfield  {journal} {\bibinfo
  {journal} {Comp. Phys. Comm.}\ }\textbf {\bibinfo {volume} {183}},\ \bibinfo
  {pages} {1760} (\bibinfo {year} {2012})}\BibitemShut {NoStop}%
\bibitem [{\citenamefont {Johansson}\ \emph {et~al.}(2013)\citenamefont
  {Johansson}, \citenamefont {Nation},\ and\ \citenamefont
  {Nori}}]{Johansson2013}%
  \BibitemOpen
  \bibfield  {author} {\bibinfo {author} {\bibfnamefont {J.~R.}\ \bibnamefont
  {Johansson}}, \bibinfo {author} {\bibfnamefont {P.~D.}\ \bibnamefont
  {Nation}}, \ and\ \bibinfo {author} {\bibfnamefont {F.}~\bibnamefont
  {Nori}},\ }\bibfield  {title} {\enquote {\bibinfo {title} {Qutip 2: A python
  framework for the dynamics of open quantum systems},}\ }\href {\doibase
  10.1016/j.cpc.2012.11.019} {\bibfield  {journal} {\bibinfo  {journal} {Comp.
  Phys. Comm.}\ }\textbf {\bibinfo {volume} {184}},\ \bibinfo {pages} {1234}
  (\bibinfo {year} {2013})}\BibitemShut {NoStop}%
\end{thebibliography}
\end{document}


\draft
\title{Supplementary Materials for ``Manipulating complex hybrid entanglement and testing multipartite Bell inequalities in a superconducting circuit"}

\author{Y.~Ma}
\thanks{These two authors contributed equally to this work.}
\affiliation{Center for Quantum Information, Institute for Interdisciplinary Information Sciences, Tsinghua University, Beijing 100084, China}

\author{X.~Pan}
\thanks{These two authors contributed equally to this work.}
\affiliation{Center for Quantum Information, Institute for Interdisciplinary Information Sciences, Tsinghua University, Beijing 100084, China}

\author {W.~Cai}
\affiliation{Center for Quantum Information, Institute for Interdisciplinary Information Sciences, Tsinghua University, Beijing 100084, China}

\author {X.~Mu}
\affiliation{Center for Quantum Information, Institute for Interdisciplinary Information Sciences, Tsinghua University, Beijing 100084, China}

\author {Y.~Xu}
\affiliation{Center for Quantum Information, Institute for Interdisciplinary Information Sciences, Tsinghua University, Beijing 100084, China}

\author {L.~Hu}
\affiliation{Center for Quantum Information, Institute for Interdisciplinary Information Sciences, Tsinghua University, Beijing 100084, China}

\author {W.~Wang}
\affiliation{Center for Quantum Information, Institute for Interdisciplinary Information Sciences, Tsinghua University, Beijing 100084, China}

\author {H.~Wang}
\affiliation{Center for Quantum Information, Institute for Interdisciplinary Information Sciences, Tsinghua University, Beijing 100084, China}

\author {Y.~P.~Song}
\affiliation{Center for Quantum Information, Institute for Interdisciplinary Information Sciences, Tsinghua University, Beijing 100084, China}

\author{Zhen-Biao~Yang}
\thanks{E-mail:zbyang@fzu.edu.cn}
\affiliation{Fujian Key Laboratory of Quantum Information and Quantum Optics,\\
College of Physics and Information Engineering, Fuzhou University, Fuzhou,\\
Fujian 350116, China}

\author{Shi-Biao~Zheng}
\thanks{E-mail: t96034@fzu.edu.cn}
\affiliation{Fujian Key Laboratory of Quantum Information and Quantum Optics,\\
College of Physics and Information Engineering, Fuzhou University, Fuzhou,\\
Fujian 350116, China}

\author{L.~Sun}
\thanks{Email: luyansun@tsinghua.edu.cn}
\affiliation{Center for Quantum Information, Institute for Interdisciplinary Information Sciences, Tsinghua University, Beijing 100084, China}

\maketitle

\section{Experimental Device and Parameters}
\begin{figure*}
\includegraphics{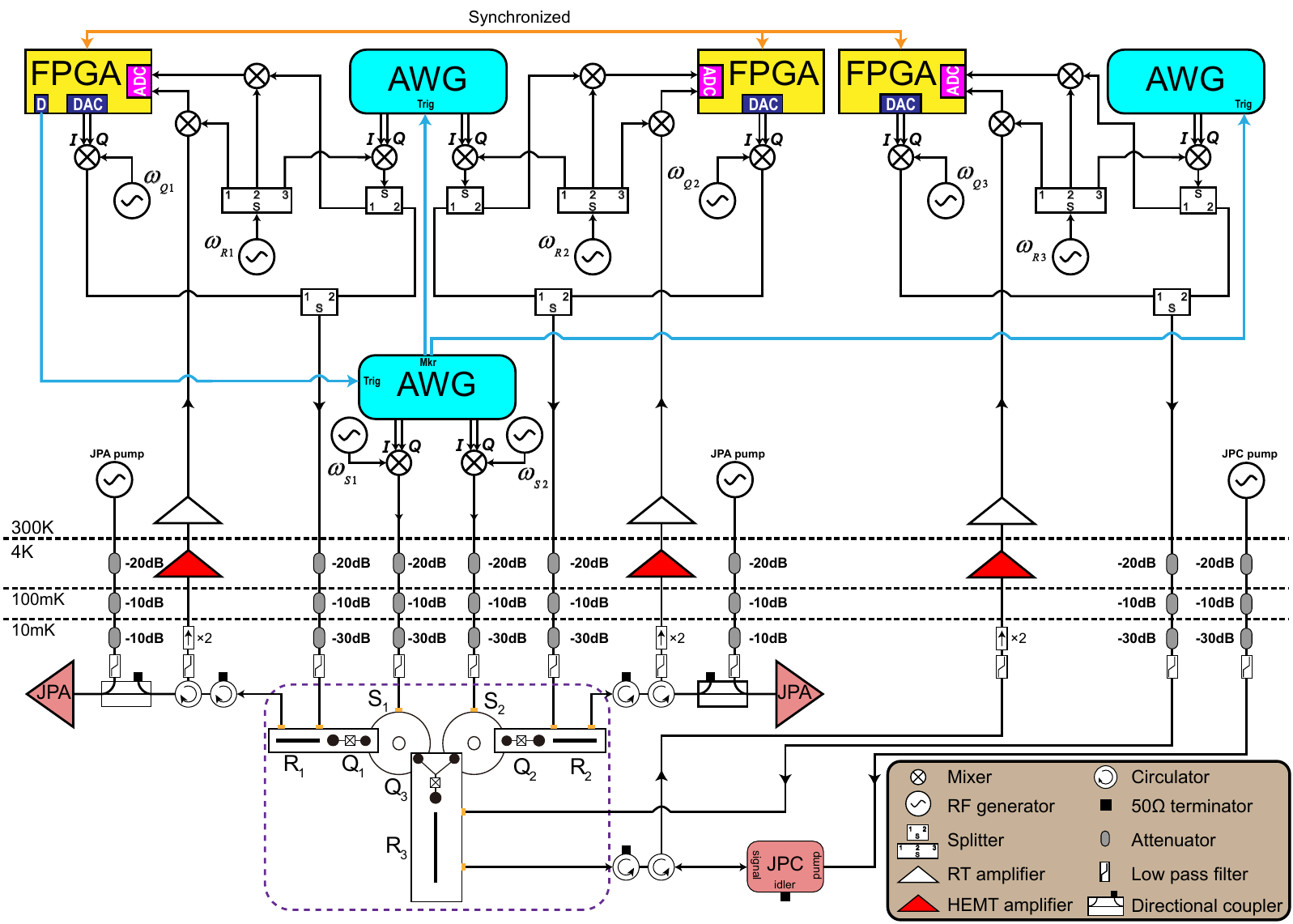}
\centering
\caption{\textbf{Experimental setup.} Detailed wiring diagram of the experimental setup.}
\label{fig:setup}
\end{figure*}

The experimental device consists of two coaxial cavities ($S_1$, $S_2$) as the storage cavities~\cite{Reagor2016-s}, dispersively coupled to three superconducting transmon qubits ($Q_1$, $Q_2$, $Q_3$). The device is machined with high-purity aluminum and wet etched for high surface quality. The three qubits and their corresponding stripline readout cavities ($R_1$, $R_2$, $R_3$)~\cite{Axline2016-s} are fabricated separately on three sapphire chips and assembled together with mechanical clamping. Quantum limited amplifiers~\cite{Vijay2009-s,Bergeal2010Nature-s} are used for high-fidelity single-shot readout of the qubits. We use two separate Josephson parametric amplifiers (JPA) for $Q_1$ and $Q_2$, and a Josephson parametric converter (JPC) for $Q_3$. The detailed device geometry can be found in Ref.~\cite{Xu2019-s}. The measurement setup is shown in Fig.~\ref{fig:setup}, similar to that in Ref.~\cite{Ma2019-s}. We use three field programmable gate arrays (FPGA) to control the three qubits. However, in Ref.~\cite{Ma2019-s} only one storage cavity, one transmon qubit, and one readout cavity are used.

Due to the nonlinearities of qubits, the self-Kerr effect $-\frac{1}{2}\chi_{ii}a_i^\dagger{a}_i^\dagger{a_ia_i}$ ($i=1,2$) of each storage cavity should also be considered in the full system Hamiltonian for a better control and optimization of the experiment. Here $a_i^{\dagger}$ is the creation operator of the $i$-th storage cavity. We list all the main parameters relevant to the experiment in Tables~\ref{param1} and \ref{param2}.

\begin{figure*}
\includegraphics{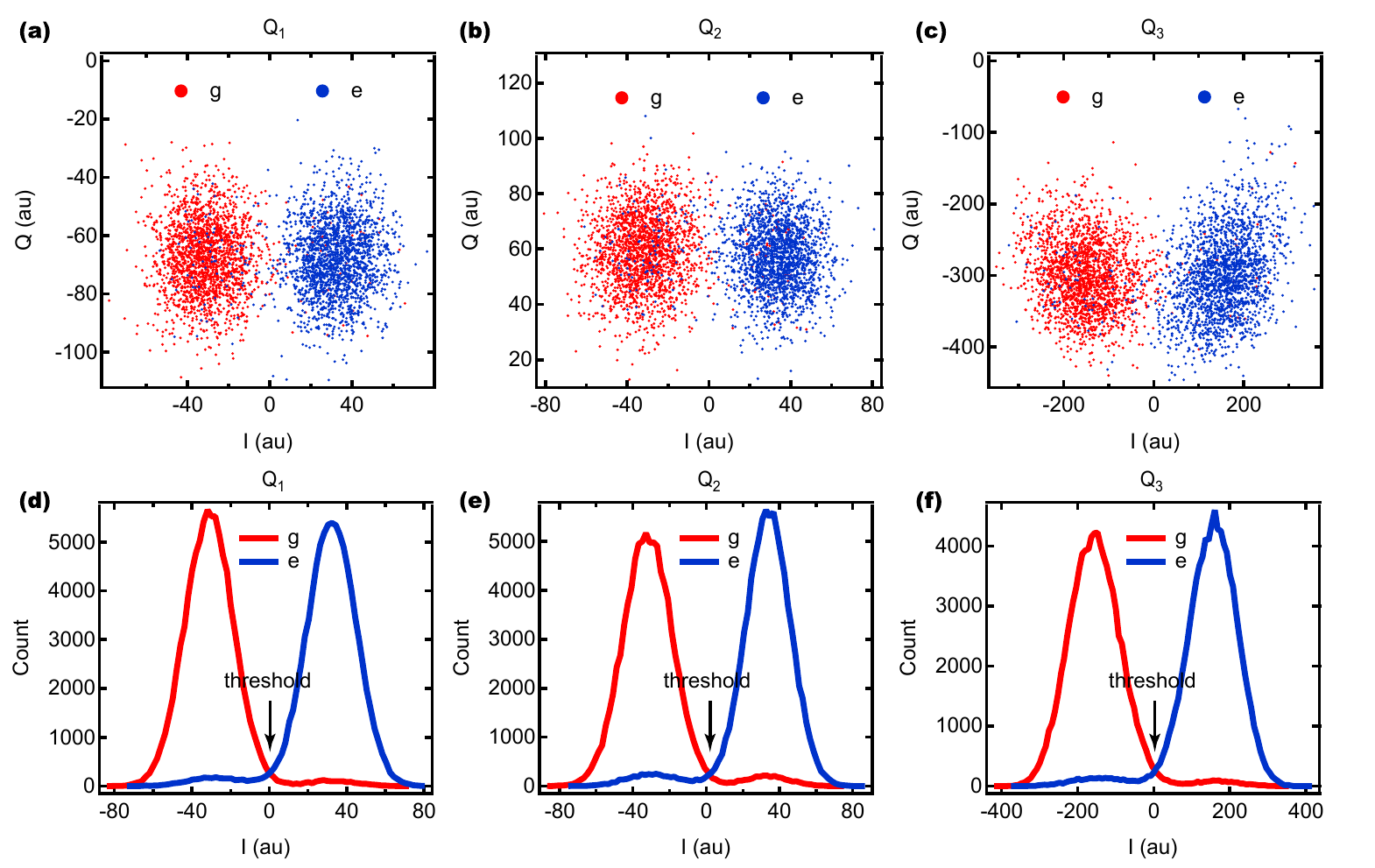}
\centering
\caption{\textbf{Qubit readouts.} (a-c) Readout results in the $I-Q$ complex plane. (d-f) Histograms of the individual qubit readouts.}
\label{fig:histograms}
\end{figure*}

\renewcommand\arraystretch{1.3}
\begin{table*}\scriptsize
\centering
\begin{tabular}{p{2cm}<{\centering}c p{2cm}<{\centering}c p{2cm}<{\centering}c
 p{2cm}<{\centering}c p{2cm}<{\centering}c p{2cm}<{\centering}c
 p{2cm}<{\centering}c p{2cm}<{\centering}c p{2cm}<{\centering}c}
\hline
\multirow{2}*{Mode} & \multicolumn{8}{c}{Nonlinear terms: $\chi_{ij}/2\pi$(MHz)}\\
& $Q_1$ & $R_1$ & $S_1$ & $Q_2$ & $R_2$ & $S_2$ & $Q_3$ & $R_3$\\ \hline
$Q_1$ & 252  & $2.0$ & $1.6$ & 0.005 & --- & --- & 0.032 & --- \\
$S_1$ & $1.6$ & --- & 0.005   & --- & --- & 0.004 & $0.524$ & ---\\
$Q_2$ & 0.005 & --- & --- & 207 & $2.1$ & $2.67$ & 0.067 & ---\\
$S_2$ & --- & --- & 0.004 & $2.67$ & --- & 0.016 & 1.494 & ---\\
$Q_3$ & 0.032 & --- & $0.524$ & 0.067 & --- & 1.494 & 151 & 1.54 \\
\hline
\end{tabular}
\caption{\textbf{Measured nonlinear coupling terms including both cross Kerr and self-Kerr effects.}}
\label{param1}
\end{table*}

\begin{table}\scriptsize
\centering
\begin{tabular}{p{2cm}<{\centering}c p{2cm}<{\centering}c p{2cm}<{\centering}c p{2cm}<{\centering}c}
\hline
Mode & Frequency (GHz) & $T_1$ ($\mu$s) & $T_{2}^*$ ($\mu$s)\\ \hline
$Q_1$ & 6.036 & 35 & 17\\
$R_1$ & 8.892 & 0.058 & ---\\
$S_1$ & 6.594 & 600 & 500\\
$Q_2$ & 5.170 & 20 & 12\\
$R_2$ & 8.800 & 0.055 & ---\\
$S_2$ & 6.050 & 700 & 290\\
$Q_3$ & 5.560 & 20 & 25\\
$R_3$ & 9.032 & 0.086 & ---\\
\hline
\end{tabular}
\caption{\textbf{Frequencies and coherence properties of the device.}}
\label{param2}
\end{table}

\section{Estimation of measured Bell signal visibility}
\label{visibility}
Readout property is critical for the measurement of Bell signal. To estimate the reduced visibility of Bell signal due to imperfect detections, we first calibrate the fidelities of the quantum non-demolition (QND) readouts of the qubits and parity measurements of the cavities. The experimental results are summarized in Tables~\ref{T:readoutFidelity1} and \ref{T:readoutFidelity2}.

\subsection{Calibration of qubit readout fidelity}

For each qubit readout, we calibrate the QND readout fidelity ($p_{gg},p_{ee}$) and state distinguishability ($p_M$) separately. The calibration results are presented in Table~\ref{T:readoutFidelity1}. The QND readout fidelity $p_{gg}$ ($p_{ee}$) is the probability of obtaining the same result of the ground state $\ket{g}$ (excited state $\ket{e}$) provided the initial state is $\ket{g}$ ($\ket{e}$). These fidelities are mainly limited by qubit damping and thermal effect. In experiment, we calibrate the individual QND readout fidelity by preparing a superposition state of each qubit followed by two consecutive readouts. The fidelity $p_{gg}$ ($p_{ee}$) is extracted by post-selecting $\ket{g}$ ($\ket{e}$) state in the first readout and counting the probability of getting the same result in the second readout.

The qubit state distinguishability $p_M$ is related to the separation of the two different readout results for $\ket{g}$ and $\ket{e}$, and is determined by the signal-to-noise ratio of the readout. All the readout histograms of the qubits have been shown in Fig.~\ref{fig:histograms}. Each histogram is fitted with two Gaussian functions with the same standard deviation and an appropriate threshold in the middle is chosen to digitize the readout signal to +1 and -1 for $\ket{g}$ and $\ket{e}$, respectively. $p_M$ is then calculated as the proportion of the Gaussian distribution that is correctly addressed by the threshold. Since the threshold is nearly in the exact middle of two separate Gaussian distributions, $p_M$ is approximately equal for both $\ket{g}$ and $\ket{e}$.


\subsection{Calibration of parity measurement fidelity}
\label{parity_pulse}
In the experiment, we use parity measurements after certain displacements in the storage cavity to extract the expectation values of the Pauli operators on the photonic states. The parity measurement is achieved by sandwiching a conditional cavity $\pi$-phase shift between two $R_y^{\pi/2}$ rotations (protocol $0-0$). Due to the cross-Kerr between the readout and storage cavities and the finite bandwidth of the $R_y^{\pi/2}$ pulses, the large photon-number states in our experiment can lead to systematic errors on the parity measurement. To overcome this problem, we perform another round of parity measurement, in which the rotation axis of the second $\pi/2$ pulse $R_{-y}^{\pi/2}$ is changed to $-y$ (protocol $0-\pi$), and average the two parity measurement results as the final detected parity~\cite{Vlastakis2013-s}.

To obtain the parity measurement fidelity, we prepare a coherent state $\ket{\alpha=1.5}$ and then run six rounds of parity measurements. The first five parity measurements are with the proper and identical protocols, such that we can post-select cases with the first five measurement results all ending up with $\ket{g}$ and indicating the same parity as the ideally even or odd parity state. The final 6-th parity measurement results are to obtain the corresponding parity measurement fidelities, which are summarized in Table~\ref{T:readoutFidelity2}. It can be checked that $0\text{-}0$ ($0\text{-}\pi$) protocol has a bit higher fidelity for odd (even) parity measurement as expected because the qubit ends up with the $\ket{g}$ state.


\begin{table} [t]
\centering
     \begin{tabular*}{0.48\textwidth}{@{\extracolsep{\fill}}@{\extracolsep{\fill}}cccc}
        \hline
        Qubit readout fidelity & $p_{gg}$ & $p_{ee}$ & $p_M$  \\\hline
        $Q_1$ & $0.989$ & $0.943$ & $0.994$ \\
        $Q_2$ & $0.986$ & $0.925$ & $0.991$ \\
        $Q_3$ & $0.985$ & $0.914$ & $0.991$ \\
        \hline
    \end{tabular*} \vspace{-6pt}
    \caption{\textbf{Fidelities of the qubit readouts.} $p_{gg}$ ($p_{ee}$) is the probability of obtaining the same result of $\ket{g}$ ($\ket{e}$) provided the initial state is $\ket{g}$ ($\ket{e}$). The lower value of $p_{ee}$ mainly comes from the qubit decoherence during the waiting time to dissipate readout photons after projecting the qubit to the initial $\ket{e}$ state and the following measurement time. $p_M$ is the proportion that is correctly addressed by the threshold in the readout histogram.}
    \label{T:readoutFidelity1}
\end{table}
\begin{table}
\begin{tabular*}{0.48\textwidth}{@{\extracolsep{\fill}}@{\extracolsep{\fill}}ccc}
        \hline
        $S_1$ protocol & even & odd  \\\hline
        $0-\pi$ & $0.979$ & $ 0.958$ \\
        $0-0$ & $0.963$ & $ 0.973$ \\
        \hline
        \hline
        $S_2$ protocol & even & odd  \\\hline
        $0-\pi$ & $0.965$ & $ 0.951$ \\
        $0-0$ & $0.953$ & $ 0.976$ \\
        \hline
    \end{tabular*} \vspace{-6pt}
    \caption{\textbf{Fidelities of the parity measurements.} The numbers correspond to the probabilities of measuring even (odd) parity provided the initial parity state is ideally even (odd). The loss of fidelity mainly comes from the qubit decoherence between the two $\pi/2$ rotations.}
    \label{T:readoutFidelity2}
\end{table}

\subsection{Estimation of measured Bell signal visibility}
Here we estimate the reduced visibility of the Bell signal. When we measure the expectation values of the Pauli operators on the GHZ state, we have equal probabilities of measuring $\ket{g}$ and $\ket{e}$ (even and odd) states. Therefore, we simplify the QND readout fidelity as $p_{\mathrm{QND}}=(p_{gg}+p_{ee})/2$. Similarly, the parity measurement fidelity $p_S$ for each storage cavity can be represented as the average of the four values in Table~\ref{T:readoutFidelity2}. The qubit readout fidelity $p_Q$ can be considered as a combination of $p_M$ and the qubit damping. Taking into account the average damping effect at half of the readout time of $T_r=600$~ns, the qubit readout fidelity $p_Q$ can then be approximated as:
\begin{equation}
p_Q=p_M(1+e^{-\frac{T_r}{2T_1}})/2.
\end{equation}

Here we consider the probabilities of two cases where no error or only one error happens during the joint Wigner and Bell measurement process [Fig.~1(c) in the main text]. The probability of having no error can be estimated as:
\begin{equation}
P_0=p_{Q3}.p_{\mathrm{QND1}}.p_{\mathrm{QND2}}.p_{S1}.p_{S2}=0.845,
\end{equation}
where the numbers in subscripts represent the labelling of the transmon qubit or the storage cavity. Similarly, the probability of having one error can be estimated as:
\begin{equation}
\begin{aligned}
P_1=&(1-p_{Q3}).p_{\mathrm{QND1}}.p_{\mathrm{QND2}}.p_{S1}.p_{S2} \\
&+p_{Q3}.(1-p_{\mathrm{QND1}}).p_{\mathrm{QND2}}.p_{S1}.p_{S2} \\
&+p_{Q3}.p_{\mathrm{QND1}}.(1-p_{\mathrm{QND2}}).p_{S1}.p_{S2} \\
&+p_{Q3}.p_{\mathrm{QND1}}.p_{\mathrm{QND2}}.(1-p_{S1}).p_{S2} \\
&+p_{Q3}.p_{\mathrm{QND1}}.p_{\mathrm{QND2}}.p_{S1}.(1-p_{S2}) \\
=&0.145.
\end{aligned}
\end{equation}
Here one error on $Q_1$ ($Q_2$) means that an error occurs after the first qubit measurement but before the following parity measurement (a finite waiting time to dissipate the readout photons), approximately corresponding to $p_{\mathrm{QND1}}$ ($p_{\mathrm{QND2}}$). Each of these errors leads to a sign flip of the Bell signal. Neglecting higher-order errors, we can get the visibility of the Bell signal:
\begin{equation}
\mathscr{V}=P_0-P_1=0.700.
\end{equation}

This reduced visibility agrees well with the experimental result ($8.381/12.29=0.682$) presented in the main text and also indicates that the decoherence has little effect on the Bell state generation because of the short total generation time.


\section{Experimental Sequences and Numerical Optimization}
\noindent{}\textbf{Initialization.} Before generating the Bell state, we apply  individual measurements on all three qubits, followed by standard parity measurements on both storage cavities. We ensure the initial system state is $\ket{g}_1\ket{g}_2\ket{g}_3\ket{0}_1\ket{0}_2$ by post-selecting the correct measurement results. This step has been neglected in Fig.~1(c) of the main text.

\noindent{}\textbf{Correlation measurements.} All the qubit and cavity states have to be measured. This is done by first measuring the qubit and then also using the known qubit state to infer the following cavity parity measurement result. Therefore, we record every measurement result of $Q_1$ and $Q_2$. The parity measurement results are averaged over two rounds with both $0-0$ and $0-\pi$ protocols as explained above.


%
%
%

\noindent{}\textbf{Cavity reset.} In order to reduce experiment repetition time when performing Wigner tomography, we reset the storage cavity by using a four-wave mixing technique between the readout and the storage cavities~\cite{Pfaff2017-s}, which results in a fast leakage of photons in the storage cavity. When measuring the Bell signal, however, we use the standard natural cooling method with sufficient waiting time for a better reset.

\begin{figure}
\includegraphics{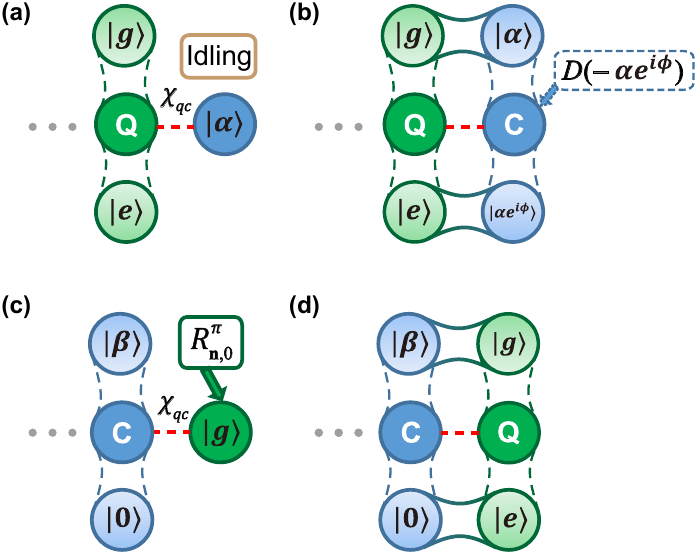}
\centering
\caption{\textbf{Sequential preparation of larger GHZ states.} After preparing an entangled chain comprised of alternatively distributed qubits and cavities, we can extend the entangled chain in two cases. (a-b) Entangle an extra cavity to an existed qubit. (a) We first displace the new cavity to a coherent state $|\alpha\rangle$ with sufficiently large $|\alpha|$. (b) Then a conditional phase shift on the cavity through the $\chi_{qc}$ term (the dispersive coupling between the qubit and the cavity) leads to different coherent states depending on the qubit states and realizes the entanglement. One component of cavity needs to be displaced to the vacuum state if one wants to connect and entangle another extra qubit to this cavity. (c-d) Entangle an extra qubit to an existed cavity. (c) The initial two-fold state of the cavity should contain a vacuum component and a coherent state component. We apply the vacuum-dependent rotation $R_{\textbf{n},0}^{\pi}$ on the qubit to complete the entanglement. (d) Final state of the system with an extra entangled qubit.}
\label{generalization}
\end{figure}

\noindent{}\textbf{Larger GHZ states.} The number of the entangled items in the GHZ state can be easily increased in a sequential way as shown in Fig.~\ref{generalization}. To entangle an extra cavity that couples to the existed qubit, we first displace the new cavity to a coherent state $|\alpha\rangle$ with sufficiently large amplitude $|\alpha|$ and then perform a conditional phase shift on the cavity through the dispersive coupling between the qubit and the new cavity. To entangle an extra qubit that couples to the existed cavity, we can apply a vacuum-dependent qubit rotation on the qubit to complete the entanglement.

\begin{figure}
\includegraphics{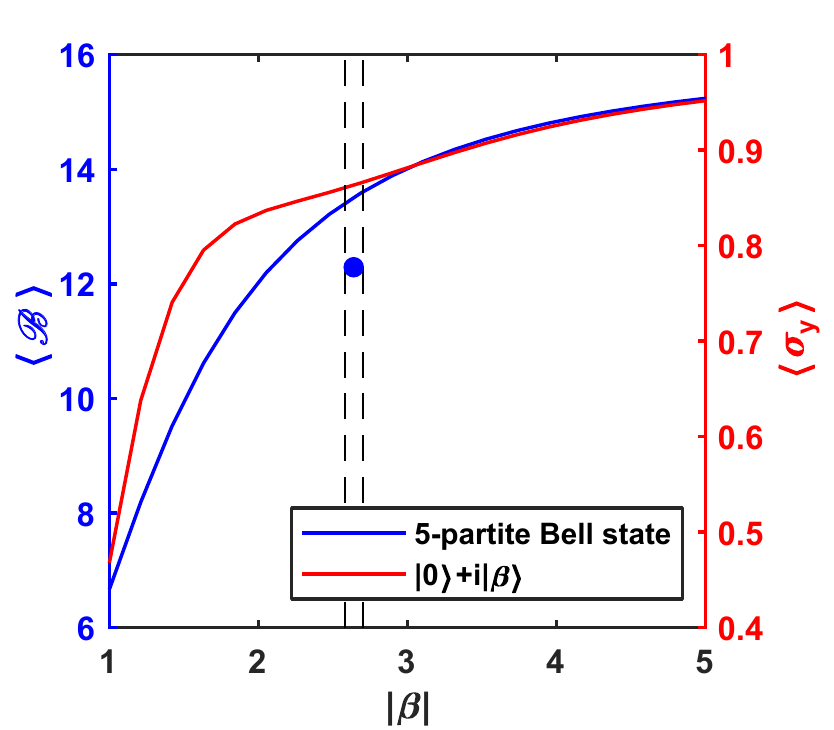}
\centering
\caption{\textbf{Bell signal vs the amplitude of the coherent states.} The blue curve is the expectation value of the Bell operator for the ideal GHZ state in the form of Eq.~(4) in the main text with $\theta=0$ and $\beta_1=\beta_2=\beta$. The red curve is the expectation value of $\sigma_y$ operator on a single cavity in the general superposition state $|0\rangle+i|\beta\rangle$. We mark the experimental parameters $|\beta_1|=2.58,|\beta_2|=2.71$, as the two reference vertical lines; the Bell signal $\langle\mathscr{B}\rangle=12.29$ with a filled circle, obtained based on the simulated GHZ state under these parameters. The gap between the simulated Bell signal (filled circle) and the value for the corresponding ideal GHZ state (blue line) is mainly caused by the deformation from Kerr effects in the GHZ state generation.}
\label{swp-alpha}
\end{figure}

\begin{figure}
\includegraphics{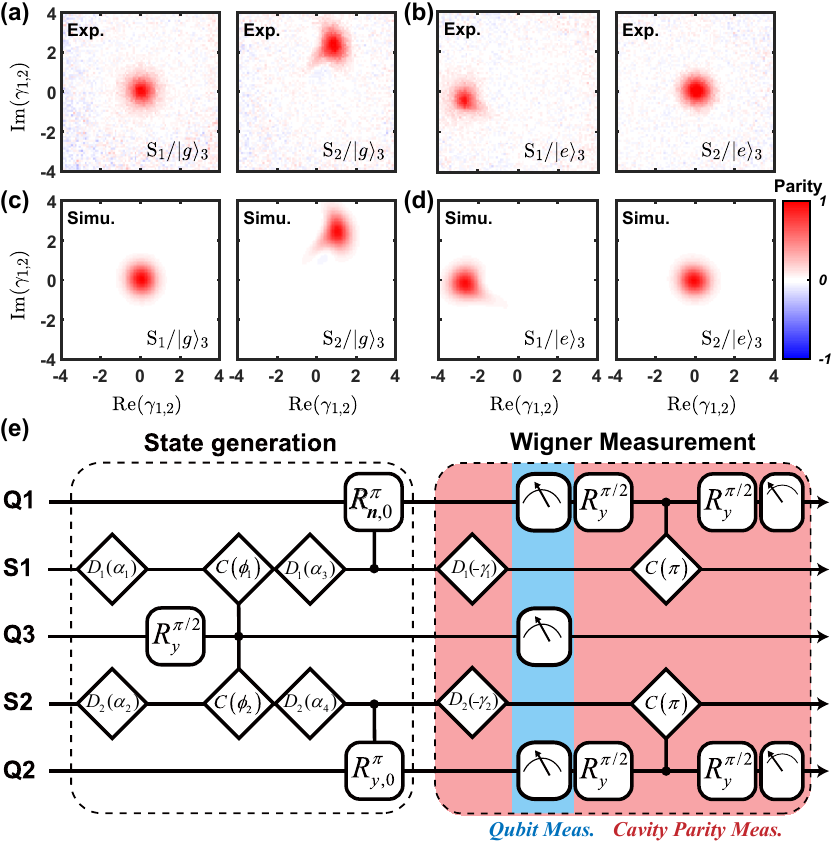}
\centering
\caption{\textbf{Qubit-state-dependent Wigner tomography.} (a) Measured single-cavity Wigner functions of the two cavities with $Q_3$ being projected to the ground state $\ket{g}$. The parity peak center of $S_1$ is located at the center of $\text{Re}(\gamma)\text{-Im}(\gamma)$ plane as expected. The deformed Wigner function of $S_2$ is mainly caused by the self-Kerr effect. (b) Measured single-cavity Wigner functions but with $Q_3$ being projected to the excited state $\ket{e}$. The parity peak of $S_2$ is now located at center and the rather weak self-Kerr effect of $S_1$ causes small deformation of $S_1$'s Winger function. (c-d) Simulation results with $Q_3$ being projected to $\ket{g}$ and $\ket{e}$, respectively. (e) Experimental sequence for the state generation and Wigner measurement, similar to Fig.~1(c) of the main text. We note that since the rotation axis of the conditional qubit rotation on $Q_1$ is fixed as \textbf{n}, we omit the variation parameter $\theta$.}
\label{wigner}
\end{figure}

\noindent{}\textbf{Numerical optimization.} We determine the pulse parameters with the exact experimental sequence [Fig.~1(c) of the main text] by a numerical optimization. The variables to be optimized are ``$\alpha_1=\alpha_2=\mathrm{real~number}$" and the duration $\tau$ of the conditional phase pulses on the storage cavities. Our optimization target is to maximize the expectation value of the Bell operator with respect to the evolved final state. In the numerical simulation for generating the GHZ state, we neglect the decoherence effects because of the restriction of the large computational space. However, we include the self-Kerr effects and the cross Kerr effect between the two storage cavities that are listed in Table~\ref{param1} but not considered in the theoretical part of the main text. Since the Kerr effects during the single qubit and cavity operations deviate the state from the ideal one, we take these Kerr effects into consideration and apply corrections to the parameters $\alpha_3$ and $\alpha_4$. In order to calculate the expectation value of the Bell operator, we extract the values of $\beta_1$ and $\beta_2$ from the simulated results of the qubit-state-dependent Wigner tomographies, as discussed in Section~\ref{sd-wigner}.

\indent{}The optimization results can be understood as a trade-off between the approximate definition of the cavity Pauli $\sigma_y$ operator and the systematic operation errors for large photon number states. As shown in Fig.~\ref{swp-alpha}, the expectation value of Pauli $\sigma_y$ operator for a single cavity (red curve) and the Bell signal for the ideal 5-partite GHZ state in the form of Eq.~(4) in the main text (blue curve) approach the theoretical upper bounds of 1 and 16, respectively, as the amplitude of the coherent states increases. On the other hand, the coherent states with larger amplitude would suffer from faster deformation of the states because of Kerr effects. As a result, in the optimization based on the numerical simulation, we finally choose coherent states with medium amplitudes (black dashed lines). Correspondingly, the Bell signal $\langle\mathscr{B}\rangle=12.29$ (filled blue circle) is obtained based on the simulated GHZ state without including the system decoherence in the GHZ state generation and the detection imperfections in the subsequent Bell measurement (Section~\ref{visibility}).

\section{Qubit-State-Dependent Wigner Function}
\label{sd-wigner}
The experiment to measure the conditional single-cavity Wigner function of the prepared GHZ state is displayed in Fig.~\ref{wigner}. The prepared state for both the experiment and simulation is the same as the one used in Fig.~2 of the main text. The experimental data of both cavities conditioned on different $Q_3$ states are shown in Fig.~\ref{wigner}(a-b), while the simulation results are shown in Fig.~\ref{wigner}(c-d). The simulation data are obtained by first applying a projection operator on $Q_3$ to the simulated final state and then partially tracing the other subsystems for Wigner function calculation. We take the peak locations of $S_1$'s Wigner function in Fig.~\ref{wigner}(d) and $S_2$'s Wigner function in Fig.~\ref{wigner}(c) for the parameters $\beta_1$ and $\beta_2$ in our experiment, respectively.

\indent{}The experimental sequence for the state generation and Wigner measurement is shown in Fig.~\ref{wigner}(e). This sequence is similar to Fig.~1(c) in the main text except that there are no pre-rotations of the qubits when performing the qubit measurements. The rotation axis of the conditional qubit rotation on $Q_1$ is fixed as \textbf{n}. We implement the Wigner tomographies on the two cavities simultaneously to save time since the two parity measurements commute. We apply the method mentioned in Section~\ref{parity_pulse} to obtain the final parity data.


%